\documentclass[sigconf]{acmart}

\usepackage{subcaption}  
\usepackage{multirow}  
\usepackage{xspace}
\usepackage{hyperref}

\AtBeginDocument{%
  \providecommand\BibTeX{{%
    \normalfont B\kern-0.5em{\scshape i\kern-0.25em b}\kern-0.8em\TeX}}}

\copyrightyear{2021}
\acmYear{2021}
\setcopyright{acmcopyright}
\acmConference[SIGIR '21]{Proceedings of the 44th International ACM SIGIR Conference on Research and Development in Information Retrieval}{July 11--15, 2021}{Virtual Event, Canada}
\acmBooktitle{Proceedings of the 44th International ACM SIGIR Conference on Research and Development in Information Retrieval (SIGIR '21), July 11--15, 2021, Virtual Event, Canada}
\acmPrice{15.00}
\acmISBN{978-1-4503-8037-9/21/07}
\acmDOI{10.1145/3404835.3462958}

\begin{document}
\fancyhead{}

\title{Unsupervised Proxy Selection for Session-based Recommender Systems}

\author{Junsu Cho}
\affiliation{%
  \institution{Pohang University of Science and Technology}
  \city{Pohang}
  \country{South Korea}}
\email{junsu7463@postech.ac.kr}

\author{SeongKu Kang}
\affiliation{%
  \institution{Pohang University of Science and Technology}
  \city{Pohang}
  \country{South Korea}}
\email{seongku@postech.ac.kr}

\author{Dongmin Hyun}
\affiliation{%
  \institution{Pohang University of Science and Technology}
  \city{Pohang}
  \country{South Korea}}
\email{dm.hyun@postech.ac.kr}

\author{Hwanjo Yu}
\authornote{Corresponding Author}
\affiliation{%
  \institution{Pohang University of Science and Technology}
  \city{Pohang}
  \country{South Korea}}
\email{hwanjoyu@postech.ac.kr}

% %%
% %% By default, the full list of authors will be used in the page
% %% headers. Often, this list is too long, and will overlap
% %% other information printed in the page headers. This command allows
% %% the author to define a more concise list
% %% of authors' names for this purpose.

\newcommand{\proposed}{ProxySR\xspace}

\begin{abstract}
Session-based Recommender Systems (SRSs) have been actively developed to recommend the next item of an anonymous short item sequence (i.e., session).
Unlike sequence-aware recommender systems where the whole interaction sequence of each user can be used to model both the \emph{short-term interest} and the \emph{general interest} of the user, the absence of user-dependent information in SRSs makes it difficult to directly derive the user's general interest from data.
Therefore, existing SRSs have focused on how to effectively model the information about short-term interest within the sessions, but they are insufficient to capture the general interest of users.
To this end, we propose a novel framework to overcome the limitation of SRSs, named \proposed, which imitates the missing information in SRSs (i.e., general interest of users) by modeling \emph{proxies} of sessions. 
\proposed \emph{selects a proxy} for the input session in an unsupervised manner, and combines it with the encoded short-term interest of the session.
As a proxy is jointly learned with the short-term interest and selected by multiple sessions, a proxy learns to play the role of the general interest of a user and \proposed learns how to select a suitable proxy for an input session. 
Moreover, we propose another real-world situation of SRSs where a few users are logged-in and leave their identifiers in sessions, and a revision of \proposed for the situation. 
Our experiments on real-world datasets show that \proposed considerably outperforms the state-of-the-art competitors, and the proxies successfully imitate the general interest of the users without any user-dependent information.
\end{abstract}

\begin{CCSXML}
<ccs2012>
   <concept>
       <concept_id>10002951.10003317.10003347.10003350</concept_id>
       <concept_desc>Information systems~Recommender systems</concept_desc>
       <concept_significance>500</concept_significance>
       </concept>
   <concept>
       <concept_id>10002951.10003260.10003261.10003269</concept_id>
       <concept_desc>Information systems~Collaborative filtering</concept_desc>
       <concept_significance>500</concept_significance>
       </concept>
   <concept>
       <concept_id>10002951.10003260.10003261.10003271</concept_id>
       <concept_desc>Information systems~Personalization</concept_desc>
       <concept_significance>500</concept_significance>
       </concept>
   <concept>
       <concept_id>10010147.10010257.10010282.10010292</concept_id>
       <concept_desc>Computing methodologies~Learning from implicit feedback</concept_desc>
       <concept_significance>500</concept_significance>
       </concept>
 </ccs2012>
\end{CCSXML}

\ccsdesc[500]{Information systems~Recommender systems}
\ccsdesc[500]{Information systems~Collaborative filtering}
\ccsdesc[500]{Information systems~Personalization}
\ccsdesc[500]{Computing methodologies~Learning from implicit feedback}

\keywords{Session-based Recommender System, Collaborative Filtering, Proxy}


\maketitle

\section{Introduction}
% 1. Session-based Recommender System
In the era of information explosion, Recommender Systems (RSs) play critical roles in providing users with interesting contents in many online applications such as e-commerce or music application. 
Most conventional RSs discover the users' preferences based on their long-term interaction history with items, and then predict the next item of interest based on their preferences \cite{mf, bpr, caser, fpmc}.
However, most users (i.e., users without logging in) tend not to leave their profiles while browsing online services \cite{srgnn, fgnn, csrm, gcegnn}.
That is, in numerous real-world online services, what RSs can refer to are the short-term (e.g., in a day) sequences of item consumption (i.e., \emph{sessions}) left by anonymous users.
Accordingly, Session-based Recommender Systems (SRSs) have been actively developed to provide more accurate recommendations for the next items of the sessions without any user-dependent information. 

\begin{figure}[t]
    \centering
    \begin{subfigure}[b]{0.45\textwidth}
        \centering
        \includegraphics[width=\textwidth, trim={0.0cm 0.0cm 0cm 0.0cm}, clip] {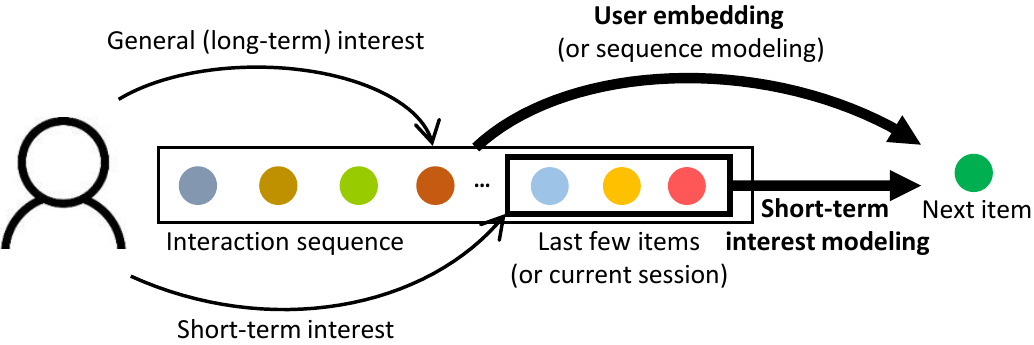}
        \caption{Sequence-aware RS}
        \label{fig:intro_seq}
    \end{subfigure}
    \hfill
    \begin{subfigure}[b]{0.45\textwidth}
        \centering
        \includegraphics[width=\textwidth, trim={0.0cm 0.3cm 0cm 0.6cm}, clip] {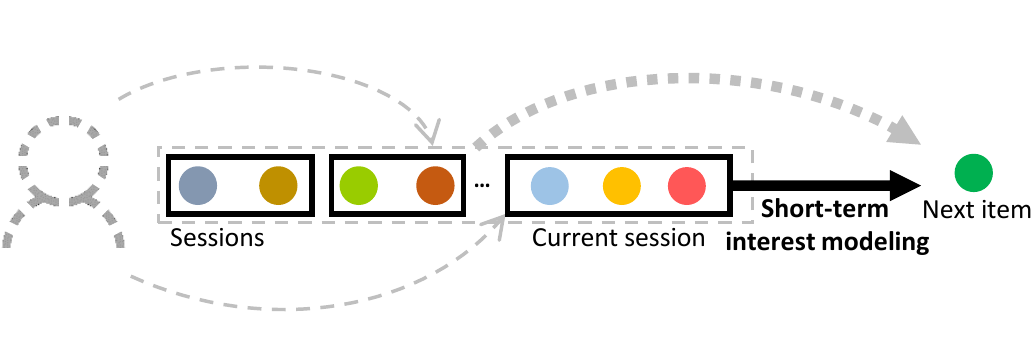}
        \caption{SRS}
        \label{fig:intro_sess}
    \end{subfigure}
    \vspace{-0.2cm}
    \caption{Difference between the amount of information available in sequence-aware RSs and SRSs. The gray dotted lines indicate the unavailable information in SRSs.}
    \vspace{-0.4cm}
\end{figure}

% 2. Challenges in SRS
The absence of the user-dependent information causes a challenge for the SRSs.
With the user-dependent information, sequence-aware RSs which utilize the whole interaction sequence of each user can model a user's \emph{general (or long-term) interest} via a sequence modeling or a user embedding \cite{sasrec, attrec}, in addition to the \emph{short-term (or current) interest} within the last few interactions (Fig. \ref{fig:intro_seq}). 
The general interest of a user is the user's individual preference which changes slowly over time, and discovering it increases the recommendation performance as it influences the user's next behavior along with the user's short-term interest \cite{generalinterest1, generalinterest2, generalinterest3}.
However, SRSs, only using the anonymous sessions, capture only the short-term interest within the sessions and have a limitation in directly deriving the user's general interest (Fig. \ref{fig:intro_sess}).
Although many SRSs have been developed recently, this limitation has not been addressed. 

Most existing methods focus on how to effectively extract useful information from a single session \cite{gru4rec, narm, stamp, srgnn, fgnn}, and thus cannot consider the relationships between sessions. 
To address this problem, some recent methods define the relationships using the item co-occurrence between the sessions and achieve the state-of-the-art recommendation performances \cite{csrm, gcegnn}.
However, they do not consider the relationships between sessions that are more complex than the item co-occurrence.
Several studies pointed out that the general interest of users is too complex to model only with relationships based on the item co-occurrence \cite{hybridsvd, magnn}.
Therefore, the existing methods that employ a static rule based on the item co-occurrence to define the relationships between sessions are insufficient to capture the general interest of users. 

% 4. propose
To overcome the limitation of the SRSs, we propose a novel SRS framework named \proposed, which imitates the missing information (i.e., general interest of users) by modeling \emph{proxies}, each of which encodes information shared by multiple sessions.
\proposed is designed inspired by the following characteristics of general interest: 1) multiple sessions have similar general interests in them (i.e., multiple sessions are created by a user who has a steady general interest), and 2) the general interest can compensate for the missing information in SRSs other than the short-term interest in predicting the next item.  
Specifically, in addition to the session encoder capturing the short-term interest within the input session, \proposed includes a separate component that \emph{selects a proxy} from a set of proxies in an unsupervised manner.
The encoded short-term interest and the selected proxy are combined, and the two modeling components are jointly learned in an end-to-end manner to accurately predict the next item of the input session. 
Through the joint learning, \proposed learns how to select a suitable proxy to fill in the missing information other than the short-term interest in predicting the next item, and at the same time, the proxy selected by several sessions learns the information common to those sessions. 

Moreover, we establish another real-world situation of SRSs where a few users are logged-in and leave their identifiers in sessions, and propose a revision of \proposed for the situation. 
In this situation, \proposed can assign more suitable proxies using the user identifiers to provide more accurate predictions.
Our extensive experiments on real-world datasets show that \proposed considerably outperforms the state-of-the-art competitors.
Our analyses also show that the proxies actually imitate the general interest of users without any user-dependent information, and play an important role in predicting the next item of sessions.
Lastly, we show that \proposed provides more accurate recommendations as the ratio of known users increases.

\vspace{-0.2cm}
\section{Related Work}

\subsection{Session-based Recommender Systems}
SRSs aim at predicting the next item of each session. Without any user-dependent information, the only information that SRSs can utilize is the chronologically-ordered item sequence in each session which implies the short-term interest of user. Accordingly, some existing methods focus on how to effectively modeling the information in each single session. For example, GRU4Rec \cite{gru4rec} uses GRU \cite{gru} which takes the embeddings of items in a session as input, to model the sequential patterns in the session. NARM \cite{narm} summarizes the hidden states of GRU using an attention module, to model the user's main purpose and sequential patterns in the session. STAMP \cite{stamp} incorporates each item information in a session according to its similarity to the last item based on an attention mechanism, to focus on the most recent interest. SASRec \cite{sasrec} uses a self-attention network to capture the user's preference within a sequence. SR-GNN \cite{srgnn}, which is the first attempt to express the sessions in directed graphs, captures the complex transitions of items in a session via graph neural networks. FGNN \cite{fgnn} introduces an attentional layer and a new readout function in graph neural networks to consider the latent order rather than the chronological item order in a session. RepeatNet \cite{repeatnet} first predicts whether the next item will be a repeat consumption or a new item, and then predicts the next item for each case. GRec \cite{grec} leverages future data as well when learning the preferences for target items in a session for richer information in dilated convolutional neural networks.

However, these methods cannot consider the relationships between sessions, as they use only the information within a single session. 
To overcome this limitation, some recent methods define the relationships between sessions using the item co-occurrence between them. 
CSRM \cite{csrm} incorporates information of the latest few sessions according to their similarity to the current session. 
CoSAN \cite{cosan} extends CSRM to find out the similar sessions for each item, not for each session. GCE-GNN \cite{gcegnn}, which shows the state-of-the-art recommendation performance, constructs a global graph that models pairwise item-transitions over sessions. 
However, all these approaches do not consider the general interest of users, which is important to increase the recommendation performance but too complex to be captured only by the relationships based on the item co-occurrence between sessions \cite{hybridsvd, magnn}. 

\begin{figure*}[t]
    \centering
    \includegraphics[width=1.5\columnwidth, trim={1.0cm 0.4cm 1.6cm 1.2cm}, clip]{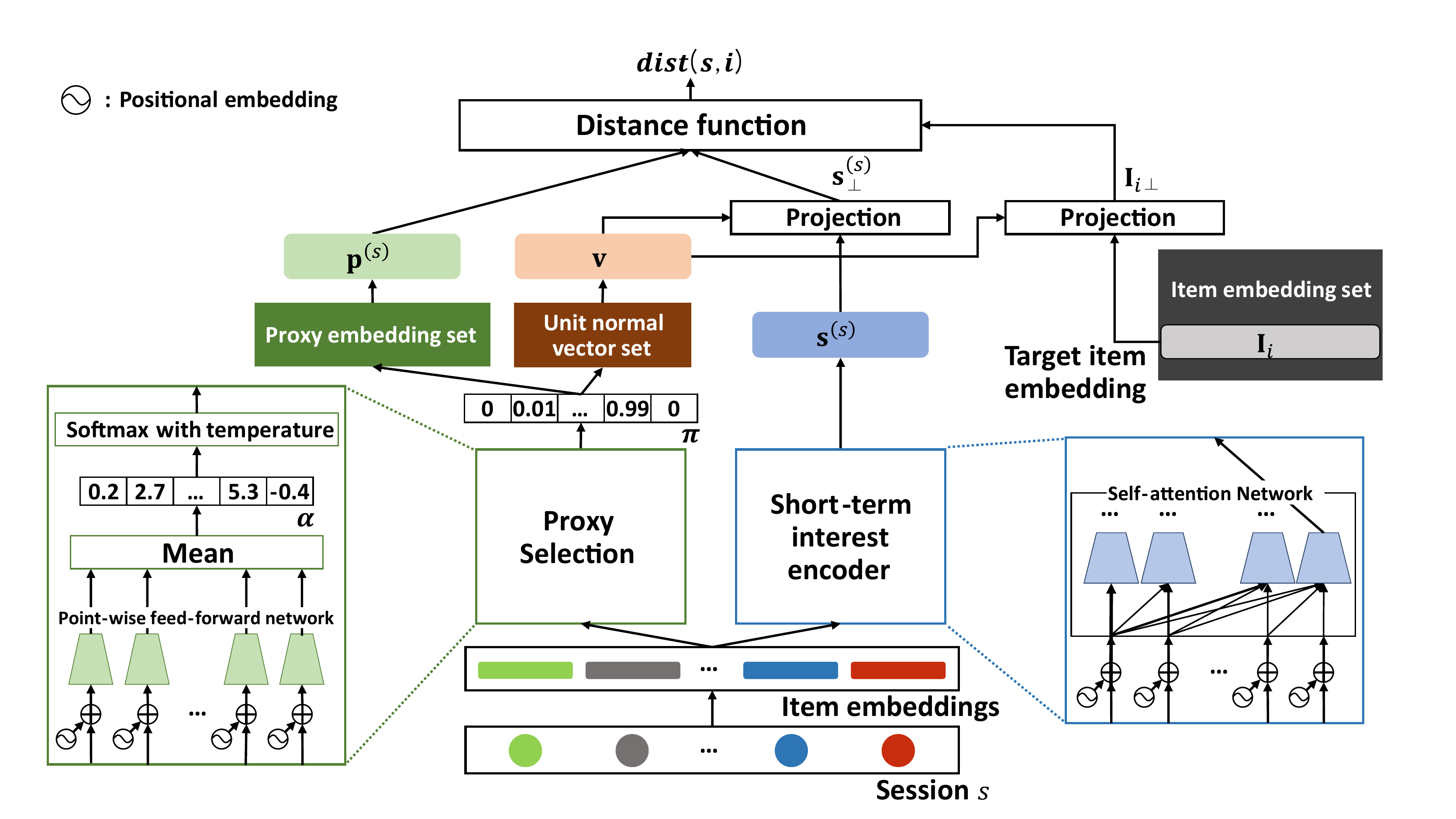}
    \vspace{-0.2cm}
    \caption{The overall architecture of \proposed.}
    \vspace{-0.2cm}
    \label{fig:model}
\end{figure*}

\vspace{-0.2cm}
\subsection{Learning with Proxies}
Recently, there have been many attempts to learn the model using proxies, each of which is a representative of several instances, in order to revise a conventional operation between the instances. 
For example, in computer vision, Proxy-NCA \cite{proxynca} firstly employed the proxy triplets in metric learning instead of the triplets of data instances, which reduces the number of triplets and improves the robustness of the model against noisy labels. 
SoftTriple \cite{softtriple} assigns multiple proxies for an image class, as a class may contain multiple local clusters due to the intra-class variance. 
Although it is adopted to various methods, the main role of a proxy in them is to learn common information about a set of data instances. 

Likewise, a proxy in \proposed models the information common to several sessions and serves as a representative of them. 
A distinctive characteristic of the proxy in \proposed is that it imitates the general interest of a user, by being shared across several sessions and used in combination with the short-term interest. 
As a result, \proposed provides more accurate recommendations by alleviating the problem of SRSs mentioned above.

\vspace{-0.1cm}
\section{Method}

This section first introduces the task of SRSs and the notation in this paper (Section \ref{section:problemformulation}), then describes the details of \proposed (Fig. \ref{fig:model}). 
\proposed selects a proxy for the input session (Section \ref{section:proxyselection}) and encodes the session into a short-term interest representation (Section \ref{section:shortterminterestencoder}), and then uses the aggregation of them to define the distance function between the session and the target item (Section \ref{section:combination}). 
Finally, the loss function for training \proposed is proposed using the distance function (Section \ref{section:training}). Moreover, we establish another real-world scenario for SRSs, and propose a revised version of \proposed for the scenario (Section \ref{section:realistic}).

\subsection{Problem Formulation and Notation}
\label{section:problemformulation}
In this paper, we aim to recommend the next item for an input session.
Let $\mathbf{I} \in \mathbb{R}^{N \times d}$ denote the item embedding matrix where $N$ is the number of items and $d$ is the embedding size. 
Given a session $s = [s_1, s_2, ..., s_n]$, where $s_* \in \{1, 2, ..., N\}$ is the index of an item in session $s$, $n$ is the number of items in $s$ (i.e., $n=|s|$), and the interactions are chronologically-ordered, our goal is to recommend top-$k$ items as the next item $s_{n+1}$. 
In the training phase, the model is trained to predict every item $s_t$ ($t \le n$) in $s$ using $[s_1, s_2, ..., s_{t-1}]$. 

Throughout this paper, we use a bold capital letter for a matrix (e.g., $\mathbf{I}$, $\mathbf{P}$), a bold small letter for a vector (e.g., $\mathbf{p}$, $\boldsymbol{\pi}$), a superscript $P$ for a modeling component for the proxy selection (e.g., $\mathbf{W}^{P,(1)}$, $\mathbf{E}^{P}$), and a superscript $S$ for a modeling component for the short-term interest encoder (e.g., $\mathbf{W}^{S,(1)}$, $\mathbf{E}^{S}$). Also we denote the $i$-th vector (or element) of a matrix (or vector) as a subscript $i$ (e.g., $\mathbf{I}_{s_j}$, $\boldsymbol{\pi}_j$).

\subsection{Proxy Selection}
\label{section:proxyselection}
This section describes how \proposed obtains a proxy $\mathbf{p}^{(s)}$ for an input session $s$. 
Given an input session, \proposed selects a proxy from the predefined set of proxies in an unsupervised manner, and combines it with the encoded short-term interest to make the final representation of the session. 
Through the end-to-end learning, the modeling component for proxy selection learns how to select a suitable proxy for an input session, and the selected proxy learns the information common to the sessions that select the proxy. As the proxy fills in the missing information other than the short-term interest in predicting the next item, the proxy imitates the general interest of the user. 

To this end, \proposed first uses the input session $s$ to build a skewed probability distribution to select a proxy embedding from a predefined set of $K$ proxy embeddings. More specifically, \proposed utilizes an encoder network to produce logits of the probabilities, and then converts them to the skewed probability distribution $\boldsymbol{\pi} \in \mathbb{R}^K$ using a softmax function with a temperature parameter \cite{temperature} as follows: 
\begin{equation}
\begin{split}
    \boldsymbol{\alpha} &=  f^P(s) \\
    \boldsymbol{\pi}_{j} &= {\frac{\exp({\boldsymbol{\alpha}_j / \tau})} {\sum_{j'=1}^K \exp({\boldsymbol{\alpha}_{j'} / \tau})}} \text{ for } j \in \{1, 2, ..., K\}
\end{split}
\label{equation:alpha}
\end{equation}
where $f^P$ is an encoder network for a session where $f^P(s) \in \mathbb{R}^{K}$, $K$ is the predefined number of proxies, $\boldsymbol{\pi}_{j}$ is the probability for the $j$-th proxy, and $\tau > 0$ is the temperature parameter. As $\tau$ gets smaller, $\boldsymbol{\pi}$ becomes a hard distribution close to a one-hot vector, and as $\tau$ gets larger, $\boldsymbol{\pi}$ becomes a uniform distribution where every element is close to $1/K$. Therefore, we assign a large initial value to $\tau$ and decrease it as the training progresses, because if the initial value of $\tau$ is small, the gradients are skewed to few logits, which is not desirable for the unstable initial training \cite{gumbel, temperature}. Finally, we obtain the proxy embedding $\mathbf{p}^{(s)}$ for session $s$ as follows:
\begin{equation}
\begin{split}
    \gamma &= {\frac{\sum_{j=1}^K \left\lVert \boldsymbol{\pi}_{j}  \mathbf{P}_j \right\rVert_2} {\left\lVert \sum_{j=1}^K \boldsymbol{\pi}_{j} \mathbf{P}_j \right\rVert_2}} \\
    \mathbf{p}^{(s)} &= \gamma \cdot \sum_{j=1}^K \boldsymbol{\pi}_{j} \mathbf{P}_j
\end{split}
\label{equation:pi1}
\end{equation}
where $\mathbf{P} \in \mathbb{R}^{K \times d}$ is the proxy embedding matrix. When $\tau$ is small enough after several training epochs, $\boldsymbol{\pi}$ becomes a one-hot vector and only one proxy embedding is selected from the set. 

When $\tau$ is large in the initial training phase, the scale of obtained proxy can be too small because each of the proxies is randomly initialized with a mean of zero and is uniformly aggregated to cancel out each other. 
Therefore, we prevent this problem by rescaling the obtained proxy with $\gamma$ which forces its $l_2$-norm to maintain the weighted mean of the $l_2$-norms of proxies. 

\subsubsection{Implementation Details}
Any encoder network for a sequence that captures the sequential pattern in the input sequence can be used as $f^P$. In our experiments, as a non-linear network for sessions of any length, we use two-layer point-wise feed-forward networks for the item embeddings in a session and take the average of the outputs as the logits. Also we add a learnable positional embedding \cite{bert, positionalembedding} to each item embedding, which encodes information about its position, in order to model the sequential pattern. That is, the encoder network for proxy selection in our experiments is built as follows:
\begin{equation}
    f^P(s)={\frac{1} {n}} \sum_{j=1}^{n} {\mathbf{W}^{P,(2)}}^\top \sigma \left( {\mathbf{W}^{P,(1)}}^\top \left(\mathbf{I}_{s_j} + \mathbf{E}^P_j \right) \right)
\end{equation} 
where $\mathbf{E}^P_j \in \mathbb{R}^{d}$ is the learnable positional embedding for the $j$-th position, and $\mathbf{W}^{P,(1)} \in \mathbb{R}^{d \times \lfloor(d+K)/2\rfloor}$, $\mathbf{W}^{P,(2)} \in \mathbb{R}^{\lfloor(d+K)/2\rfloor \times K}$ are the weight matrices. $\sigma$ is the Leaky ReLU activation function \cite{leakyrelu} with negative slope 0.1. 

Note that in the training phase, we employ the proxy which is selected by the whole session $s$ (i.e., $\mathbf{p}^{(s)}$) even when predicting each item $s_t$ $(t \le n)$ using $[s_1, s_2, ..., s_{t-1}]$, because each item in a session is generated with a consistent general interest of a user. Thus $\mathbf{p}^{(s)}$ learns its relationships with all items in $s$.

\subsubsection{Discussion}
To build a representation that imitates a general interest, \proposed \emph{selects} a proxy via the softmax with a temperature parameter, rather than a weighted combination of several proxies using an ordinary softmax function. 
A weighted combination of several proxies produces a unique representation for each input session, which is equivalent to encoding a session into a representation. 
However, since the general interest is difficult to be fully encoded with only the information within a single session, we cannot guarantee that the weighted combination models the general interest which is common to several sessions. 
Alternatively, \proposed models the proxies which imitate the general interests by selecting the most probable proxy, and jointly training the selected proxy with the short-term interest of the session. 
Thus, a proxy in \proposed, which is shared across the sessions that select the proxy, encodes the information common to the sessions. 
In Section \ref{section:experiments}, we provide the analyses that show the superiority of proxy selection compared to the weighted combination. 

\vspace{-0.2cm}
\subsection{Short-term Interest Encoder}
\label{section:shortterminterestencoder}
The short session itself represents the short-term (or current) interest of the user \cite{longshort2}. Therefore, \proposed encodes the input session $s$ with an encoder network into a latent representation $\mathbf{s}^{(s)}$ and uses it as the short-term interest within the session:

Specifically, we can obtain the short-term interest representation for the input session $s$ as follows:
\begin{equation}
    \vspace{-0.05cm}
    \mathbf{s}^{(s)} = f^S(s)
    \vspace{-0.05cm}
\end{equation}
where $f^S$ is a session encoder which encodes the session into a latent representation (i.e., $f^S(s) \in \mathbb{R}^{d}$), and $\mathbf{s}^{(s)} \in \mathbb{R}^{d}$ is the representation of short-term interest within session $s$.

\vspace{-0.1cm}
\subsubsection{Implementation Details}
Any session encoder can be adopted as $f^S$. In the experiments, we adopt a self-attention network \cite{transformer, sasrec} with residual connection \cite{resnet}, which effectively models a sequence considering the dependendies between the items in the sequence. 
Our short-term interest representation for the input sessions $s$ can be obtained as follows:
\begin{equation}
\begin{split}
    \mathbf{X} &= [\mathbf{I}_{s_1} + \mathbf{E}^S_{n}, \mathbf{I}_{s_2} + \mathbf{E}^S_{n-1}, ..., \mathbf{I}_{s_{n}} + \mathbf{E}^S_1]^\top \\
    \mathbf{Q} &= \text{ReLU}(\mathbf{X} \mathbf{W}^{S,(Q)})\\
    \mathbf{K} &= \text{ReLU}(\mathbf{X} \mathbf{W}^{S,(K)})\\
    \mathbf{A} &= \text{softmax}\left( \frac{\mathbf{Q} \mathbf{K}^\top} {\sqrt{d}} \right)\\
    \mathbf{Z} &= \mathbf{A} \mathbf{X} + \mathbf{X}\\
    f^S (s) &= {\mathbf{W}^{S,(2)}}^\top \text{ReLU}\left({\mathbf{W}^{S,(1)}}^\top \mathbf{Z}_n + \mathbf{b}^{S,(1)} \right) + \mathbf{b}^{S,(2)}
\end{split}
\vspace{-0.4cm}
\end{equation}
where $\mathbf{E}^S_j$ is the learnable positional embedding for the $j$-th recent interaction, $\mathbf{X} \in \mathbb{R}^{n \times d}$ is a representation of session $s$ as the input to $f^S$. $\mathbf{W}^{S,(Q)}$, $\mathbf{W}^{S,(K)}$, $\mathbf{W}^{S,(1)}$, $\mathbf{W}^{S,(2)}$ $\in$ $\mathbb{R}^{d \times d}$ are the weight matrices, and $\mathbf{b}^{S,(1)}$, $\mathbf{b}^{S,(2)}$ $\in$ $\mathbb{R}^{d}$ are the biases. Note that the positional embeddings for short-term interest encoder are assigned in reverse chronological order, to model the impact of the recent items on the short-term interest \cite{sasrec}.

\vspace{-0.2cm}
\subsection{Combination}
\label{section:combination}
\proposed adds the selected proxy and the encoded short-term interest to make the final representation of session $s$, and uses it to compute the dissimilarity score between the session and the target item $i$. 
Finally, $K$ items with the smallest dissimilarity score with $s$ are recommended. 
However, according to some precedent studies \cite{transh, transr}, a simple addition cannot model relationships within a triplet that are more complex than a one-to-one relationship. 
In other words, if the same item has to be related to two different short-term interests with the same proxy, the model forces the two short-term interests to be similar (i.e., if $\mathbf{p} + \mathbf{s}^{(1)} \approx \mathbf{I}_i$ and $\mathbf{p} + \mathbf{s}^{(2)} \approx \mathbf{I}_i$, then $\mathbf{s}^{(1)} \approx \mathbf{s}^{(2)}$). 
Likewise, if two items have to be related to the similar short-term interests with the same proxy, the model forces the two items to be similar (i.e., if $\mathbf{p} + \mathbf{s}^{(1)} \approx \mathbf{I}_{i^{(1)}}$ and $\mathbf{p} + \mathbf{s}^{(2)} \approx \mathbf{I}_{i^{(2)}}$ where $\mathbf{s}^{(1)} \approx \mathbf{s}^{(2)}$, then $\mathbf{I}_{i^{(1)}} \approx \mathbf{I}_{i^{(2)}}$). 
As SRSs may have such complex relationships, the model should be designed to capture them.

To this end, we project the encoded short-term interest and the target item embedding onto a hyperplane for the selected proxy and define the relationship between them on the hyperplane \cite{transh}. Thus, different items (or short-term interests) can have the same representation on a hyperplane, allowing \proposed to capture the complex relationships. 
Specifically, we first obtain the projected short-term interest $\mathbf{s}^{(s)}_\bot$ and the projected target item embedding ${\mathbf{I}_{i}}_\bot$ on the proxy's hyperplane as follows:
\begin{equation}
\begin{split}
    \mathbf{v} &= {\frac{\sum^K_{j=1} \boldsymbol{\pi}_j \mathbf{V}_j} {\left\lVert \sum^K_{j=1} \boldsymbol{\pi}_j \mathbf{V}_j \right\rVert_2}}\\
    \mathbf{s}^{(s)}_{\bot} &= \mathbf{s}^{(s)} - \mathbf{v}^\top \mathbf{s}^{(s)} \mathbf{v}\\
    {\mathbf{I}_i}_\bot &= \mathbf{I}_i - \mathbf{v}^\top \mathbf{I}_{i} \mathbf{v}\\
\end{split}
\label{equation:pi2}
\end{equation}
where $\mathbf{V} \in \mathbb{R}^{K \times d}$ is the unit normal vector set for the proxy hyperplanes \cite{transh}, and $\mathbf{v} \in \mathbb{R}^{d}$ is the unit normal vector for projecting onto the hyperplane for $\mathbf{p}^{(s)}$. For the normal vector to be orthogonal to the proxy's hyperplane and to have the unit length, we force $|\mathbf{v} \cdot \mathbf{p}^{(s)}| / \lVert \mathbf{p}^{(s)} \rVert_2 \le \epsilon$ and $\lVert \mathbf{V}_j \rVert_2 = 1$ with regularizers. 

Lastly, the dissimilarity score between session $s$ and target item $i$ is estimated with the distance between the projected item embedding and the aggregation of the proxy and the projected short-term interest. We compute the dissimilarity score as follows:
\begin{equation}
    dist(s, i) = \left\lVert \left( \mathbf{p}^{(s)} + \mathbf{s}^{(s)}_{\bot} \right)- {\mathbf{I}_{i}}_\bot \right\rVert^2_2
\end{equation}
As a result, a higher $dist(s, i)$ value implies a lower probability of $i$ as the next item of session $s$. 

\subsection{Training}
\label{section:training}
We use the marginal loss (i.e., hinge loss) \cite{cml, transcf} to train \proposed, so that the true next item of a session becomes closer to the session compared to the other items. In addition, we adopt the orthogonality regularizer for the unit normal vector $\mathbf{v}$ and the distance regularizer introduced in \cite{transcf}, which explicitly forces the session representation close to the target item embedding. 

Firstly, we define the loss function $\mathcal{L}$ as follows:
\begin{equation}
\begin{split}
    \mathcal{L} &= \sum_{\{s, i^+\} \in \mathcal{S}} \sum_{i^- \in NI(s)} [m + dist(s, i^+) - dist(s, i^-)]_+ \\
\end{split}
\end{equation}
where $\mathcal{S}$ is the training dataset of sessions and their true next item, $i^+$ is the true next item of session $s$, $NI(s) \subset \mathbf{I} \backslash i^+$ is a set of the negative items of $s$, $[x]_+=\max(x,0)$, and $m$ is the margin. 
Including the regularizers, we define our final objective function $\mathcal{J}$ to minimize as follows:
\begin{equation}
\begin{split}
    \text{reg}^{\text{dist}} &= \sum_{\{s, i^+\} \in \mathcal{S}} dist(s, i^+)\\
    \text{reg}^{\text{orthog}} &= \sum_{\{s, i^+\} \in \mathcal{S}} {\frac{|\mathbf{v}^{(s)} \cdot \mathbf{p}^{(s)}|} {\lVert \mathbf{p}^{(s)} \rVert_2}}\\
    \mathcal{J} &= \mathcal{L} + \lambda^{\text{dist}} \cdot \text{reg}^{\text{dist}} + \lambda^{\text{orthog}} \cdot \text{reg}^{\text{orthog}}
\end{split}
\end{equation}
where $\mathbf{v}^{(s)}$ is $\mathbf{v}$ for session $s$, and $\lambda^{\text{dist}}, \lambda^{\text{orthog}}$ are the regularization coefficients for the distance regularizer and the orthogonality regularizer, respectively. 

\vspace{-0.2cm}
\subsection{Another Real-world Case: User Semi-supervision}
\label{section:realistic}
Several studies on user behavior in the online environment report that in real world, about 50\% to 70\% of users browse the items without logging in, while the others log in and leave their user identifiers \cite{user1, user2, user3}. 
In this real-world scenario, the ground-truth user information can provide \proposed with information about which proxy to select for the input session. 
In this regard, for an input session that has its user information, we add a user bias to the logits (i.e., $\boldsymbol{\alpha}$ in Equation (\ref{equation:alpha})) for selecting a proxy, modeling the users' inclination for particular proxies. 
The other sessions without user information use the original logits. 
Thus, we only increase the probability of selecting a preferred proxy for the users, rather than forcing to assign a particular proxy according to the user information, to flexibly model which proxy the each user prefers.

Specifically, for the sessions that have their user information, $\boldsymbol{\pi}$ in Equation (\ref{equation:pi1}) and (\ref{equation:pi2}) for selecting a proxy can be replaced with $\boldsymbol{\pi}^{\text{user}}$ as follows:

\vspace{-0.4cm}
\begin{equation}
\begin{split}
    \boldsymbol{\pi}^{\text{user}}_{j}={\frac{\exp\left({(\boldsymbol{\alpha}_{j} + \mathbf{u}^{(s)}_{j}) / \tau}\right)} {\sum^K_{j'=1} \exp\left({(\boldsymbol{\alpha}_{j'} + \mathbf{u}^{(s)}_{j'}) / \tau}\right)}}\\
\end{split}
\label{equation:pi_user}
\end{equation}
where $\mathbf{u}^{(s)} \in \mathbb{R}^K$ is the learnable user bias for the user of session $s$. In Section \ref{section:experiments}, we provide the experimental results in this scenario that report the performance improvement according to the ratio of known users. 

\vspace{-0.2cm}
\section{Experiments}
\label{section:experiments}
In this section, we provide the experimental results and analyses on \proposed and other state-of-the-art baselines. Our experiments are designed to answer the following research questions. 

\begin{itemize}
    \item \textbf{RQ1}: How does \proposed perform compared to the other state-of-the-art baselines for the task of recommending the next item that have not appeared in the session?
    \item \textbf{RQ2}: How does \proposed perform for the task of next item recommendation that contains repetitive consumption? 
    \item \textbf{RQ3}: Does each component and design choice in \proposed contributes to the performance significantly? 
    \item \textbf{RQ4}: What information do the proxy embeddings encode?
    \item \textbf{RQ5}: Is the additional user information in the proposed real-world case beneficial to \proposed?
\end{itemize}

\vspace{-0.3cm}
\subsection{Experimental Settings}
\subsubsection{Datasets}
We conducted our experiments on three public datasets: \emph{RetailRocket}\footnote{https://www.kaggle.com/retailrocket/ecommerce-dataset}, \emph{Diginetica}\footnote{https://competitions.codalab.org/competitions/11161}, and \emph{LastFM}\footnote{http://ocelma.net/MusicRecommendationDataset/lastfm-1K.html} \cite{lastfm}. Table \ref{table:datasets} summarizes the statistics of each dataset. 
For datasets without the session information (i.e., RetailRocket and LastFM), we first divided each user's interactions in a day into sessions and anonymized the sessions.
Then, for all datasets, we split the sessions in each dataset into train/validation/test set in chronological order in a ratio of 8:1:1. 
Also we filtered out items in validation and test set which did not appear in the training set \cite{narm, stamp, csrm}. 

\textbf{RetailRocket} contains the users' interactions (i.e., clicks) collected from a real-world e-commerce website. This dataset is the smallest dataset we used, and has the shortest session length on average. To evaluate the models' performance on short sessions, we filtered out only the sessions that contain less than two items.

\textbf{Diginetica} has anonymous sessions of search logs from an e-commerce website. As the interactions in this dataset have their session id, we use the session ids to establish the sessions. We filtered out items that appear less than five times, and sessions with less than three items are excluded.

\textbf{LastFM} has users' music listening history. We consider an artist as an item, and used this dataset for artist recommendation \cite{repeatnet, csrm}. We filtered out items that appear less than five times, and sessions with less than three interactions or more than fifty interactions. 

\subsubsection{Evaluation}
For all the baselines, we ranked the true next item of each session in the test set among all the other items and used two widely-used metrics for ranking to evaluate the performance of top-$k$ recommendation \cite{narm, fgnn, csrm, grec}: recall (R@$k$) and mean reciprocal rank (M@$k$). We use several values for $k$: 5, 10, 20. 

\subsubsection{Baselines}
We compare the performance of \proposed with the following state-of-the-art baselines:
\begin{itemize}
    \item \textbf{GRU4Rec} \cite{gru4rec} encodes the session sequence into the final representation with GRU units. 
    \item \textbf{NARM} \cite{narm} is an RNN-based model with an attention layer that models the user's sequential behavior and main purpose.
    \item \textbf{STAMP} \cite{stamp} employs an attention mechanism to summarize a session based on the recent interest (i.e., the last item).
    \item \textbf{SASRec} \cite{sasrec} adopts a self-attention network to capture the user's preference within a sequence.
    \item \textbf{RepeatNet} \cite{repeatnet} employs a repeat-explore mechanism to predict repetitive interactions in a session.
    \item \textbf{SR-GNN} \cite{srgnn} expresses a session in a graph and encodes it with graph neural networks and an attention mechanism.
    \item \textbf{FGNN} \cite{fgnn} extends SR-GNN to model the latent order in the session with an attentional layer and a new readout function.
    \item \textbf{CSRM} \cite{csrm} utilizes memory networks to incorporate the neighbor sessions of the input session.
    \item \textbf{GRec} \cite{grec} leverages future data in a session as well when learning the preference of the session for richer information in dilated convolutional neural networks.
    \item \textbf{GCE-GNN} \cite{gcegnn} is the state-of-the-art SRS that constructs a global graph that models pairwise item-transitions over all sessions as well as the session graphs.
\end{itemize}

\begin{table}[t]
\fontsize{8}{9}\selectfont
\caption{Statistics of datasets. \# sessions is the number of sessions before the sessions are divided into sub-sessions.}
\vspace{-0.3cm}
\begin{tabular}{c|rrr}
\hline
Dataset         & \multicolumn{1}{c}{RetailRocket} & \multicolumn{1}{c}{Diginetica} & \multicolumn{1}{c}{LastFM} \\ \hline
\# interactions & 170,488                          & 713,308                        & 5,103,585                  \\
\# items        & 38,736                           & 33,950                         & 33,531                     \\
\# sessions     & 47,705                           & 101,691                        & 229,760                     \\
avg. length     & 3.57                             & 7.01                           & 22.21                      \\ \hline
\end{tabular} 
\label{table:datasets}
\vspace{-0.3cm}
\end{table}

We omitted conventional recommendation systems that require the user information (e.g., MF \cite{mf}, BPR \cite{bpr}, FPMC \cite{fpmc}, Caser \cite{caser}, and HRNN \cite{hrnn}). 

\begin{table*}[t]
\fontsize{8}{8.5}\selectfont
\setlength{\tabcolsep}{2pt}
\centering
\caption{Overall performance on the next unseen item recommendation. \emph{Imprv.} is the improvement of the performance from \proposed compared to the best performance among the other baselines. The best results are highlighted in boldface, and the second best results are underlined. }
\vspace{-0.2cm}
\begin{tabular}{c|c|cccccccccc|c|c}
\hline
Dataset                      & Metric & GRU4Rec & NARM   & STAMP  & SASRec & RepeatNet & SR-GNN & FGNN   & CSRM         & GRec   & GCE-GNN      & \proposed            & Imprv.(\%)       \\ \hline
\multirow{6}{*}{RetailRocket} & R@5    & 0.1182  & 0.1322 & 0.0780 & 0.1834 & 0.1073    & 0.1746 & 0.1308 & 0.1385 & 0.1393 & \underline{0.1876} & \textbf{0.2449} & 30.54\% \\
                              & M@5    & 0.0827  & 0.0786 & 0.0526 & 0.1145 & 0.0716    & 0.1126 & 0.0819 & 0.0815 & 0.0832 & \underline{0.1173} & \textbf{0.1454} & 23.96\% \\
                              & R@10   & 0.1513  & 0.1741 & 0.0958 & 0.2326 & 0.1289    & 0.2151 & 0.1708 & 0.1954 & 0.1852 & \underline{0.2342} & \textbf{0.3300} & 40.91\% \\
                              & M@10   & 0.0871  & 0.0842 & 0.0551 & 0.1197 & 0.0746    & 0.1180 & 0.0874 & 0.0892 & 0.0881 & \underline{0.1230} & \textbf{0.1567} & 27.40\% \\
                              & R@20   & 0.1752  & 0.2178 & 0.1092 & 0.2781 & 0.1544    & 0.2570 & 0.2061 & 0.2469 & 0.2289 & \underline{0.2787} & \textbf{0.4053} & 45.43\% \\
                              & M@20   & 0.0888  & 0.0873 & 0.0560 & 0.1245 & 0.0763    & 0.1209 & 0.0898 & 0.0928 & 0.0914 & \underline{0.1274} & \textbf{0.1622} & 27.32\% \\ \hline
\multirow{6}{*}{Diginetica}   & R@5    & 0.1124  & 0.1120 & 0.1090 & 0.1266 & 0.0959    & 0.1373 & 0.1339 & 0.1374 & 0.1313 & \underline{0.1467} & \textbf{0.1737} & 18.40\% \\
                              & M@5    & 0.0569  & 0.0565 & 0.0562 & 0.0643 & 0.0500    & 0.0712 & 0.0693 & 0.0693 & 0.0673 & \underline{0.0753} & \textbf{0.0906} & 20.32\% \\
                              & R@10   & 0.1879  & 0.1893 & 0.1808 & 0.2082 & 0.1544    & 0.2234 & 0.2164 & 0.2260 & 0.2166 & \underline{0.2321} & \textbf{0.2760} & 18.91\% \\
                              & M@10   & 0.0668  & 0.0681 & 0.0657 & 0.0750 & 0.0578    & 0.0826 & 0.0801 & 0.0809 & 0.0786 & \underline{0.0853} & \textbf{0.1041} & 22.04\% \\
                              & R@20   & 0.2963  & 0.2995 & 0.2785 & 0.3234 & 0.2326    & 0.3386 & 0.3297 & 0.3454 & 0.3321 & \underline{0.3573} & \textbf{0.4069} & 13.88\% \\
                              & M@20   & 0.0741  & 0.0753 & 0.0724 & 0.0830 & 0.0631    & 0.0905 & 0.0879 & 0.0895 & 0.0865 & \underline{0.0931} & \textbf{0.1130} & 21.37\% \\ \hline
\multirow{6}{*}{LastFM}       & R@5    & 0.0480  & 0.0455 & 0.0502 & 0.0523 & 0.0483    & 0.0564 & 0.0506 & 0.0502 & 0.0450 & \underline{0.0589} & \textbf{0.0640} & 8.66\%  \\
                              & M@5    & 0.0248  & 0.0235 & 0.0260 & 0.0282 & 0.0249    & 0.0310 & 0.0242 & 0.0255 & 0.0233 & \underline{0.0317} & \textbf{0.0335} & 5.68\%  \\
                              & R@10   & 0.0795  & 0.0752 & 0.0821 & 0.0846 & 0.0794    & 0.0910 & 0.0897 & 0.0843 & 0.0744 & \underline{0.0939} & \textbf{0.1025} & 9.16\%  \\
                              & M@10   & 0.0289  & 0.0274 & 0.0302 & 0.0324 & 0.0290    & 0.0355 & 0.0290 & 0.0300 & 0.0270 & \underline{0.0366} & \textbf{0.0385} & 5.19\%  \\
                              & R@20   & 0.1258  & 0.1207 & 0.1301 & 0.1318 & 0.1270    & 0.1423 & 0.1399 & 0.1341 & 0.1198 & \underline{0.1468} & \textbf{0.1589} & 8.24\%  \\
                              & M@20   & 0.0321  & 0.0305 & 0.0334 & 0.0356 & 0.0322    & 0.0390 & 0.0320 & 0.0334 & 0.0301 & \underline{0.0402} & \textbf{0.0424} & 5.47\%  \\ \hline
\end{tabular}
\label{table:main1}
\vspace{-0.2cm}
\end{table*}

\begin{table*}[t]
\fontsize{8}{8.5}\selectfont
\setlength{\tabcolsep}{2pt}
\centering
\caption{Overall performance on the next item recommendation with repetitive consumption. \emph{Imprv.} is the improvement of the performance from \proposed compared to the best performance among the other baselines. The best results are highlighted in boldface, and the second best results are underlined.}
\vspace{-0.2cm}
\begin{tabular}{c|c|cccccccccc|c|c}
\hline
Dataset                      & Metric & GRU4Rec & NARM   & STAMP  & SASRec & RepeatNet    & SR-GNN & FGNN   & CSRM         & GRec   & GCE-GNN      & \proposed            & Imprv.(\%) \\ \hline
\multirow{6}{*}{RetailRocket} & R@5    & 0.3840  & 0.4056 & 0.3361 & 0.3980 & 0.4357       & 0.4133 & 0.4093 & {\underline{0.4462}} & 0.4118 & 0.4346       & \textbf{0.5222} & 17.03\%    \\
                              & M@5    & 0.2797  & 0.3008 & 0.2449 & 0.2707 & 0.3105       & 0.2914 & 0.2933 & {\underline{0.3139}} & 0.3016 & 0.3108       & \textbf{0.3613} & 15.10\%    \\
                              & R@10   & 0.4341  & 0.4475 & 0.3989 & 0.4648 & 0.4540       & 0.4714 & 0.4410 & {\underline{0.5245}} & 0.4662 & 0.5114       & \textbf{0.5922} & 12.91\%    \\
                              & M@10   & 0.2866  & 0.3064 & 0.2523 & 0.2798 & 0.3143       & 0.2994 & 0.3007 & {\underline{0.3246}} & 0.3071 & 0.3178       & \textbf{0.3718} & 14.54\%    \\
                              & R@20   & 0.4758  & 0.4921 & 0.4304 & 0.5093 & 0.4697       & 0.5209 & 0.4918 & {\underline{0.5774}} & 0.5151 & 0.5582       & \textbf{0.6528} & 13.06\%    \\
                              & M@20   & 0.2894  & 0.3092 & 0.2550 & 0.2829 & 0.3157       & 0.3029 & 0.3030 & {\underline{0.3283}} & 0.3095 & 0.3211       & \textbf{0.3757} & 14.44\%    \\ \hline
\multirow{6}{*}{Diginetica}   & R@5    & 0.2064  & 0.2055 & 0.1913 & 0.2191 & 0.2335       & 0.2357 & 0.2235 & 0.2209       & 0.2302 & {\underline{0.2378}} & \textbf{0.2563} & 7.78\%     \\
                              & M@5    & 0.1163  & 0.1153 & 0.1069 & 0.1266 & 0.1308       & 0.1343 & 0.1283 & 0.1234       & 0.1342 & {\underline{0.1350}} & \textbf{0.1454} & 7.70\%     \\
                              & R@10   & 0.3037  & 0.2988 & 0.2819 & 0.3152 & 0.3137       & 0.3370 & 0.3266 & 0.3305       & 0.3324 & {\underline{0.3388}} & \textbf{0.3722} & 9.86\%     \\
                              & M@10   & 0.1292  & 0.1275 & 0.1188 & 0.1393 & 0.1415       & 0.1497 & 0.1434 & 0.1379       & 0.1487 & {\underline{0.1501}} & \textbf{0.1609} & 7.20\%     \\
                              & R@20   & 0.4167  & 0.4129 & 0.3908 & 0.4304 & 0.4016       & 0.4580 & 0.4423 & 0.4608       & 0.4571 & {\underline{0.4614}} & \textbf{0.5034} & 9.10\%     \\
                              & M@20   & 0.1370  & 0.1354 & 0.1264 & 0.1472 & 0.1476       & 0.1580 & 0.1487 & 0.1474       & 0.1569 & {\underline{0.1588}} & \textbf{0.1699} & 6.99\%     \\ \hline
\multirow{6}{*}{LastFM}       & R@5    & 0.5459  & 0.5499 & 0.5193 & 0.5502 & {\underline{0.5542}} & 0.5337 & 0.5389 & 0.5403       & 0.5251 & 0.5531       & \textbf{0.5822} & 5.05\%     \\
                              & M@5    & 0.4966  & 0.5005 & 0.4912 & 0.5042 & {\underline{0.5128}} & 0.4954 & 0.4919 & 0.5032       & 0.4936 & 0.5072       & \textbf{0.5373} & 4.78\%     \\
                              & R@10   & 0.5777  & 0.5791 & 0.5597 & 0.5796 & 0.5810       & 0.5771 & 0.5715 & 0.5689       & 0.5511 & {\underline{0.5819}} & \textbf{0.6076} & 4.42\%     \\
                              & M@10   & 0.5045  & 0.5066 & 0.5005 & 0.5108 & {\underline{0.5132}} & 0.5028 & 0.4960 & 0.5050       & 0.4970 & 0.5119       & \textbf{0.5406} & 5.34\%     \\
                              & R@20   & 0.6049  & 0.6108 & 0.5900 & 0.6120 & 0.6123       & 0.6038 & 0.5958 & 0.6025       & 0.5828 & {\underline{0.6149}} & \textbf{0.6370} & 3.59\%     \\
                              & M@20   & 0.5074  & 0.5136 & 0.5026 & 0.5147 & 0.5137       & 0.5070 & 0.5004 & 0.5093       & 0.4992 & {\underline{0.5142}} & \textbf{0.5444} & 5.77\%     \\ \hline
\end{tabular}
\label{table:main2}
\vspace{-0.2cm}
\end{table*}

\subsubsection{Implementation Details}
Each of the baselines is trained to predict every item in each session in the training dataset depending on the items before it \cite{narm, stamp, srgnn, fgnn, gcegnn}. 
Likewise, each full session instance $s=\{s_1, s_2, ..., s_{n}\}$ in the validation and test dataset is divided into its sub-sessions (i.e., $\{s_1\}$, $\{s_1, s_2\}$, $...$, $\{s_1, s_2, ..., s_{n}\}$), each with its next item \cite{narm, stamp, srgnn, fgnn, gcegnn}. 
We optimized all the baselines using Adam optimizer \cite{adam}, and tuned each hyperparameter with R@20 performance on the validation data: learning rate $\eta \in \{0.0001$, $0.0002$, $0.0005$, $0.001$, $0.002$, $0.005$, $0.01\}$, batch size $b \in \{32, 64, 128, 256, 512\}$, dropout rate \cite{dropout} $r$ $\in$ $\{0.0$, $0.1$, $0.2$, $0.3$, $0.4$, $0.5\}$, coefficient for L2 regularization $\lambda$ $\in$ $\{0.0$, $0.0001$, $0.001$, $0.01$, $0.1\}$, embedding size $d$ $\in$ $\{16$, $32$, $64$, $128\}$. Maximum length of each session is 50. We tuned the other hyperparameters of the baselines within the ranges of values provided in their papers. 
For \proposed, we bound all the embeddings within a unit sphere (e.g., $\lVert \mathbf{P}_j \rVert_2^2 \le 1$) as done in \cite{cml, transcf}. We tuned the number of proxies $K$ $\in$ $\{3$, $10$, $30$, $100$, $300$, $1000$, $3000\}$, regularization coefficient (i.e., $\lambda_{\text{dist}}$, $\lambda_{\text{orthog}}$) $\in$ $\{0.0$, $0.01$, $0.02$, $0.05$, $0.1$, $0.2$, $0.5\}$, margin $m$ $\in$ $\{0.1$, $0.2$, $0.5$, $1.0$, $2.0\}$. 
We used the exponential annealing for $\tau$: $\tau = \max(T_0 ({\frac{T_E} {T_0}})^{\frac{e} {E}}, T_E)$ where $e$ is the current training epoch, $E=10$ is the number of annealing epoch, $T_0=3$ is the initial temperature, and $T_E=0.01$ is the final temperature.

\vspace{-0.2cm}
\subsection{Task Formulation}
We conducted the experiments on two tasks: 1) next unseen item recommendation, and 2) next item recommendation with repetitive consumption. Although the previous studies \cite{narm, stamp, srgnn, fgnn, gcegnn, csrm} only focus on the second task, we claim that the task of next unseen item recommendation is more suitable for evaluating the ability of the model to capture the user's purpose within the session. 
That is because it is difficult to properly evaluate the learning ability of the model if the model can achieve a high performance by memorizing the items in the session. 
On the other hand, the next unseen item recommendation task requires a higher ability to discover the user's hidden preferences. We also claim that the next unseen item recommendation is more practical, as the repetitive consumption within a session for a short period of time is driven by the user's need for the item that the user already knows. That is, the user does not have to rely on a RS, and it is more desirable to recommend items that the user does not know. For the next unseen item recommendation in our experiments, we omitted every sub-session that contains its target item, and forced the probabilities of repetitive items to be zero when predicting the next item. Note that in our experiments, the results of all experiments and analyses except for Table \ref{table:main2} were on the task of next unseen item recommendation.

\vspace{-0.2cm}
\subsection{Performance Comparison}
\subsubsection{Overall Performance}
We measured the performance of the baselines on the test set at the epoch when the validation performance (i.e., R@20) is the best. We report the average performance from five independent runs.

Table \ref{table:main1} and Table \ref{table:main2} are the overall performance of the models on the next unseen item recommendation and the next item recommendation with the repetitive consumption for each session, respectively. We can see some notable observations from the results. 
Firstly, our proposed \proposed outperforms all the competitors on both tasks, for all the datasets. Moreover, the performance improvement of \proposed on the competitors is more significant on the next unseen item recommendation task. This result verifies the superiority of \proposed compared to the other baselines on learning the user's latent purpose within each session, as the next unseen item recommendation task requires a higher ability to discover the user's latent preferences as we claimed above.

We can also observe that the methods that utilize other information in addition to the information within the input session (i.e., CSRM, GCE-GNN, and \proposed) mostly outperform the methods that utilize only the information within the input session. This result supports our claim that a short session itself has insufficient information to fully understand the user's preferences. Moreover, the superior performance \proposed compared to CSRM and GCE-GNN proves that the information from the neighbor sessions based on the item co-occurrence is insufficient to capture the general interest of sessions. 

Furthermore, we can see that \proposed is more effective in the dataset with shorter average session length. In both tasks, the performance improvement of \proposed is the largest on RetailRocket dataset whose the average session length is the shortest, and the smallest on LastFM dataset whose the average session length is the longest. A long session may include more information than a short session, and the session itself may have information about the user's general interest rather than a short session. Therefore, \proposed which imitates the user's general interest can be more effective on the dataset with short sessions.

\begin{figure*}[t]
    \centering
    \begin{subfigure}[b]{0.26\textwidth}
        \centering
        \includegraphics[width=\textwidth, trim={0.2cm 0.3cm 2cm 1.3cm}, clip] {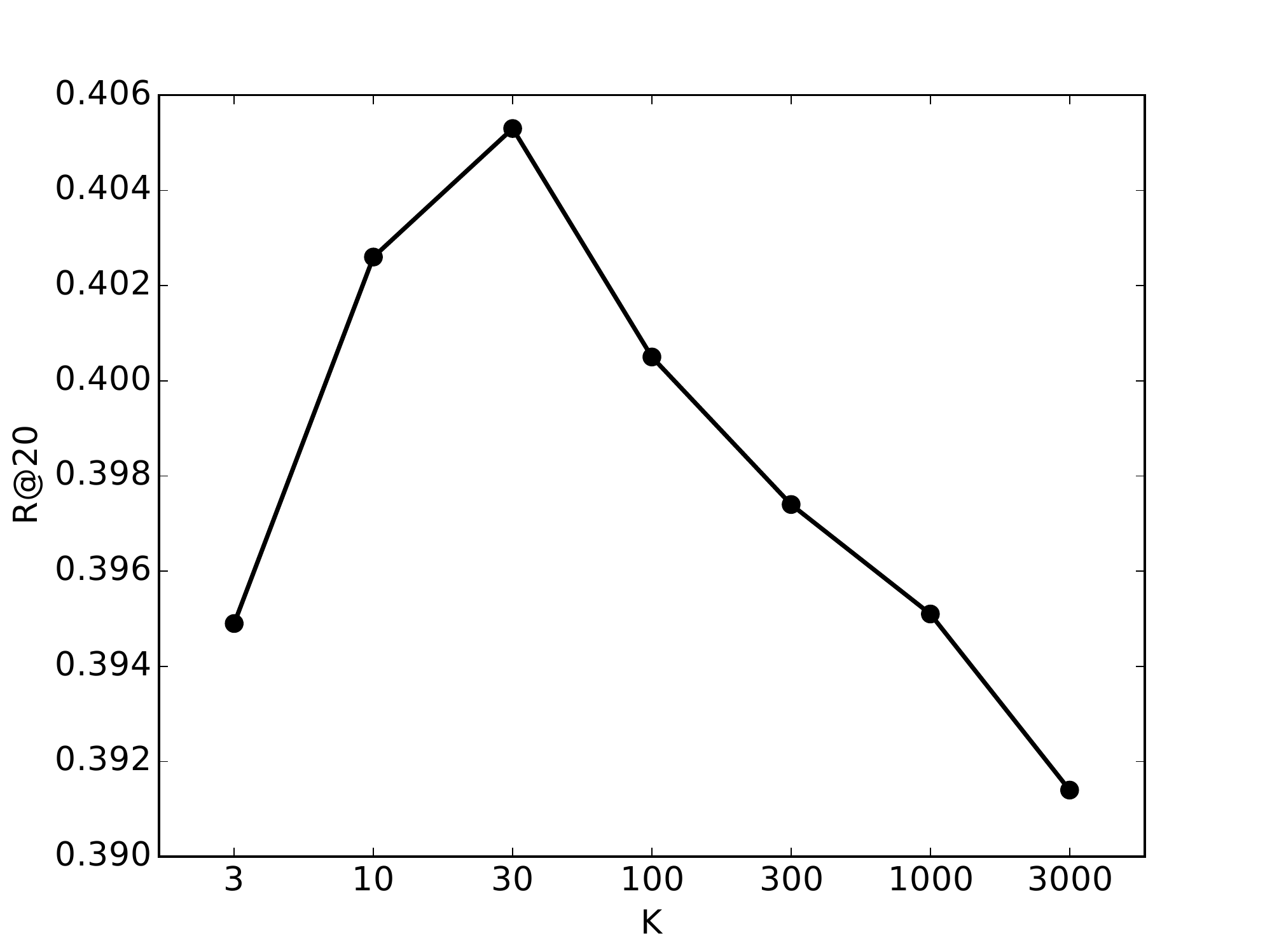}
        \caption{RetailRocket}
    \end{subfigure}
    \begin{subfigure}[b]{0.26\textwidth}
        \centering
        \includegraphics[width=\textwidth, trim={0.2cm 0.3cm 2cm 1.3cm}, clip] {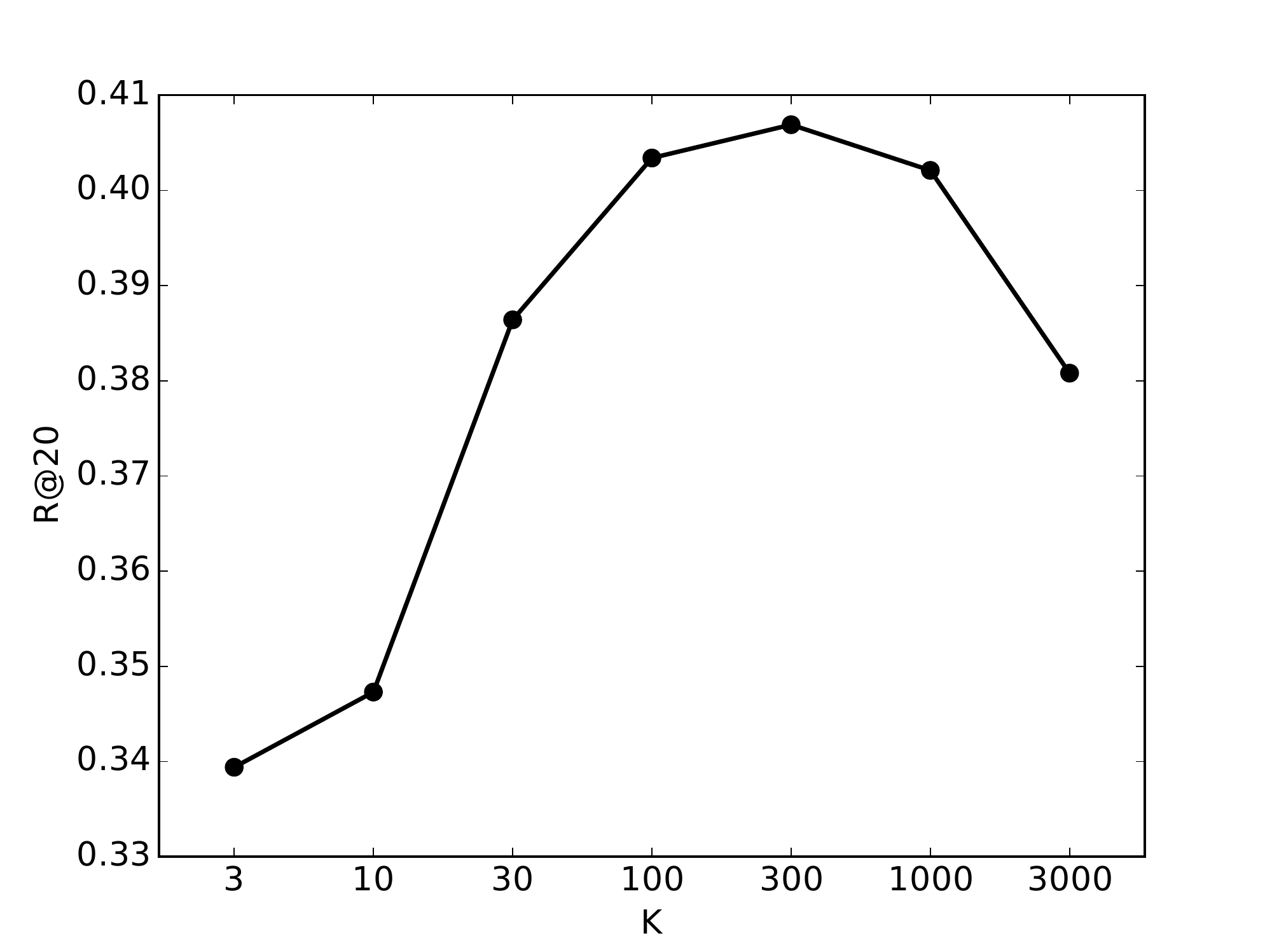}
        \caption{Diginetica}
    \end{subfigure}
    \begin{subfigure}[b]{0.26\textwidth}
        \centering
        \includegraphics[width=\textwidth, trim={0.2cm 0.3cm 2cm 1.3cm}, clip] {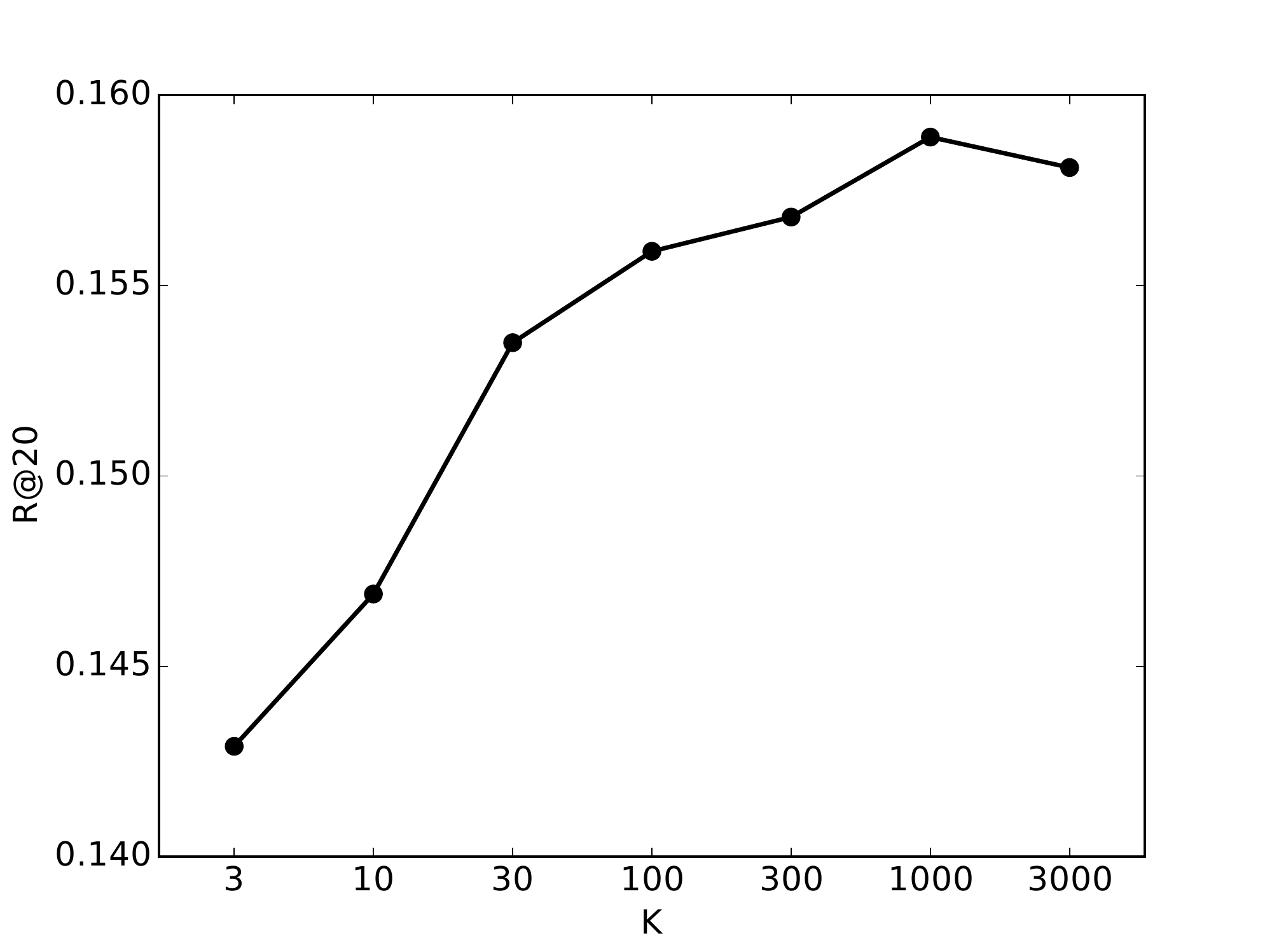}
        \caption{LastFM}
    \end{subfigure}
    \vspace{-0.2cm}
    \caption{Result of the hyperparameter parameter study on $K$ in \proposed.}
    \label{fig:hyperparameter}
    \vspace{-0.2cm}
\end{figure*}

\begin{table}[t]
\fontsize{8.9}{10}\selectfont
\centering
\caption{Performance of \proposed in the real-world scenario where a few sessions have their user information.}
\vspace{-0.2cm}
\begin{tabular}{c|cccc}
\hline
\begin{tabular}[c]{@{}c@{}}Ratio of \\ known users\end{tabular} & R@20   & M@20   & R@10   & M@10   \\ \hline
0\%                                                             & 0.1589 & 0.0424 & 0.1025 & 0.0385 \\
10\%                                                            & 0.1632 & 0.0431 & 0.1061 & 0.0391 \\
30\%                                                            & 0.1672 & 0.0443 & 0.1088 & 0.0399 \\
50\%                                                            & \textbf{0.1696} & \textbf{0.0447} & \textbf{0.1114} & \textbf{0.0402} \\ \hline
\end{tabular}
\label{table:usersemisupervision}
\vspace{-0.5cm}
\end{table}

\subsubsection{Another Real-world Case: User Semi-supervision}
We conducted an extra experiment on another real-world scenario, where a few sessions have their user information. For the experiment, \proposed uses $\boldsymbol{\pi}^\text{user}_{j}$ in Equation (\ref{equation:pi_user}) instead of $\boldsymbol{\pi}_{j}$ in Equation (\ref{equation:pi1}) and (\ref{equation:pi2}) for the sessions that have their user information.

Table \ref{table:usersemisupervision} shows the performance of \proposed according to the percentage of given user information among the users with at least 10 sessions in LastFM dataset, which has the largest number of sessions per user. The following conclusions can be drawn with the result: 1) the performance of \proposed is increased by adding the user bias to the logits for proxy selection using known user information. Moreover, as the amount of the known user information increases, \proposed makes a larger performance improvement. 2) Adding the user bias makes the logits skewed to few proxies for each known user. The improved performance by making each user have biased proxy verifies that the proxy proposed in \proposed actually encodes the user's general interest. 

\vspace{-0.2cm}
\subsection{Ablation Study}
To evaluate the impact of each component of \proposed, we provide the analysis on an ablation study with each dataset. Table \ref{table:ablation} shows the results of our ablation study. 

From the results, we can draw the following conclusions: 
1) $\mathbf{p}^{(s)}$ and $\mathbf{s}^{(s)}$ in the table, which are the results when the input session is expressed only using $\mathbf{p}^{(s)}$ and $\mathbf{s}^{(s)}$ in \proposed respectively, show worse performance than when both are used. This result suggests that both the proxy and the short-term interest are important when predicting the next item of sessions. 
2) The distance regularizer significantly improves the performance of \proposed. This verifies that it is effective to make the representation of session directly close to the next item embedding, and that it is valid to define the next item as the sum of proxy and short-term interest. 
3) \emph{No projection} is the result when the short-term interest and the target item embedding are not projected on the proxy's hyperplane (i.e., $dist(s, i) = \lVert (\mathbf{p}^{(s)} + \mathbf{s}^{(s)}) - \mathbf{I}_{i} \rVert^2_2$). As the complex relationships in SRSs are difficult to be fully modeled in a one-to-one relationship, designing \proposed to be able to capture the complex relationships improves the performance.
4) \emph{Encoding $\mathbf{p}^{(s)}$} is the result when \proposed directly encodes the proxy representation with $f^S$ instead of selecting a proxy. The result shows that for the general interest, the session information should be used only to select as the general interest is difficult to be directly derived from the session information.
5) \emph{Weighted comb.} is the result when the proxy representation is obtained by a weighted combination of multiple proxies using the ordinary softmax function instead of selecting a proxy. The result proves that, as we claimed above, selecting a proxy to let it shared across several sessions is more effective than creating a unique representation for each session. 
6) \emph{Dot product} is the result when the similarity score between the session and the target item is computed using the dot product instead of the distance function we define. Although the distance is more effective than the dot product, \proposed with the dot product still shows higher performance than other baselines due to the superiority of \proposed.

\begin{table}[t]
\fontsize{9}{9}\selectfont
\setlength{\tabcolsep}{2pt}
\centering
\caption{Result of the ablation study on each component in \proposed.}
\vspace{-0.3cm}
\begin{tabular}{l|cc|cc|cc}
\hline
                                                                 & \multicolumn{2}{c|}{RetailRocket} & \multicolumn{2}{c|}{Diginetica} & \multicolumn{2}{c}{LastFM} \\ \hline
\multicolumn{1}{l|}{}                                        
& R@20            & M@20            & R@20           & M@20           & R@20         & M@20         \\ \hline
\proposed                                         
& \textbf{0.4053}          & \textbf{0.1622}          & \textbf{0.4069}         & \textbf{0.1130}         & \textbf{0.1589}       & \textbf{0.0424}       \\ \hline
$\mathbf{p}^{(s)}$                                                            
& 0.1742          & 0.0843          & 0.1795         & 0.0362         & 0.0964       & 0.0221       \\
$\mathbf{s}^{(s)}$                                                            
& 0.3076          & 0.1244          & 0.3446         & 0.0887         & 0.1417       & 0.0385       \\
No $\text{reg}_{\text{dist}}$ 
& 0.3294          & 0.1364          & 0.2909         & 0.0816         & 0.1369       & 0.0372       \\
No projection                                                       
& 0.3202          & 0.1294          & 0.3842         & 0.1072         & 0.1492       & 0.0409       \\
Encoding $\mathbf{p}^{(s)}$
& 0.2701          & 0.1203          & 0.3209         & 0.0837         & 0.1356       & 0.0369       \\
Weighted comb.                                                       
& 0.3281          & 0.1355          & 0.3443         & 0.0953         & 0.1507       & 0.0401       \\
Dot product                                                      
& 0.3877          & 0.1571          & 0.4021         & 0.1096         & 0.1538       & 0.0416       \\ \hline
\end{tabular}
\label{table:ablation}
\vspace{-0.5cm}
\end{table}

\vspace{-0.2cm}
\subsection{Hyperparameter Study}
We performed a hyperparameter study for the number of proxy embeddings $K$ to analyse the effect of it. Figure \ref{fig:hyperparameter} shows the performances (i.e., R@20) according to the number of proxy embeddings.

For all the datasets, the performance is degraded when $K$ is too small or too large. 
If $K$ is too small, the proxies cannot be personalized for the sessions because even sessions that are less related to each other share the same proxy. 
Therefore, the proxy embeddings underfit the sessions and the model cannot provide the accurate predictions. 
On the other hand, if $K$ is too large, few sessions are allocated to each proxy, so each proxy cannot be sufficiently trained and the performance is degraded. 
Therefore, it is important to set an appropriate number of the proxies, which seems to be larger as the number of sessions in the dataset is large. 

\vspace{-0.2cm}
\subsection{Analyses on Proxies from \proposed}
\subsubsection{Information Encoded in Proxies}
This section provides an analysis on the proxies in \proposed in order to find out what information the proxies encode. 
To verify $\mathbf{p}^{(s)}$ encodes the general interest of the user of session $s$, we adopt HRNN \cite{hrnn} which is a SRS that explicitly utilizes the user information. 
HRNN trains the user embedding using the sessions of a user in sequence via a user-level RNN, and uses the user embedding as the user's general interest along with the user's next session. 
Providing each session with the user's general interest, HRNN exploits the general interest of users in addition to the sessions to enhance the accuracy of the next item prediction. 
In order to show that the proxies in \proposed actually encodes the general interest of the users, we train HRNN with the sessions grouped not by the given user information, but by the proxy they share (\emph{Proxy} in Table \ref{table:hrnn}).

Table \ref{table:hrnn} shows the overall performance of HRNN with various kinds of user information on RetailRocket dataset, where \proposed shows the largest performance improvement. 
\emph{No user} is the same as GRU4REC, and \emph{Random index} is the case when the sessions are randomly grouped. 
From the result, it can be seen that the sessions grouped by the proxy they share in \proposed also share the general interest as meaningful as the ground-truth user information. 
Even, the performance for the next item prediction is slightly higher when the sessions are grouped based on proxies rather than based on the ground-truth user information.
This result implies the proxies, which are learned in an unsupervised manner, encode information related to the general interest of users, while also containing more detailed information as needed.
As a result, through the result of significantly improving the performance compared to the results with no or incorrect user information, we can conclude that the proxies in \proposed effectively imitate the general interest of users. 

\begin{table}[t]
\fontsize{9}{9}\selectfont
\centering
\caption{Performance of HRNN with various types of the user information in it.}
\vspace{-0.3cm}
\begin{tabular}{l|cccc}
\hline
Mode         & R@20   & M@20   & R@10   & M@10   \\ \hline
No user      & 0.1738 & 0.1073 & 0.1633 & 0.1061 \\
Ground-truth & 0.1922 & 0.1157 & 0.1779 & 0.1150 \\
Proxy        & \textbf{0.1941} & \textbf{0.1187} & \textbf{0.1783} & \textbf{0.1158} \\
Random index & 0.1275 & 0.0828 & 0.1149 & 0.0804 \\ \hline
\end{tabular}
\label{table:hrnn}
\vspace{-0.5cm}
\end{table}

\vspace{-0.1cm}
\subsubsection{Visualizations}
\begin{figure}[t]
    \centering
    \begin{subfigure}[b]{0.45\columnwidth}
        \centering
        \includegraphics[width=\textwidth, trim={2.5cm 1.5cm 2cm 1.5cm}, clip] {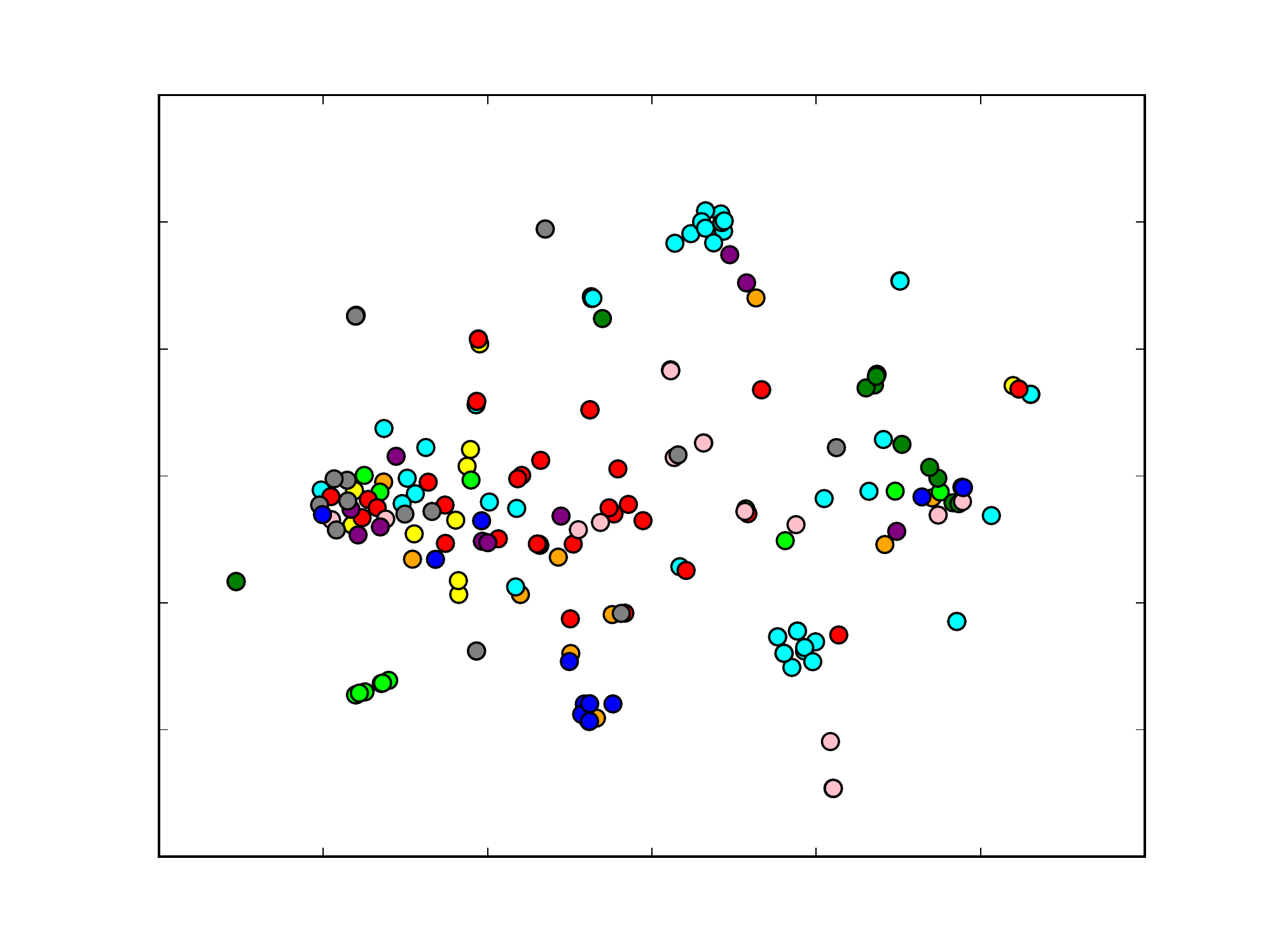}
        \caption{\proposed, 50\% trained.}
    \end{subfigure}
    \hfill
    \begin{subfigure}[b]{0.45\columnwidth}
        \centering
        \includegraphics[width=\textwidth, trim={2.5cm 1.5cm 2cm 1.5cm}, clip] {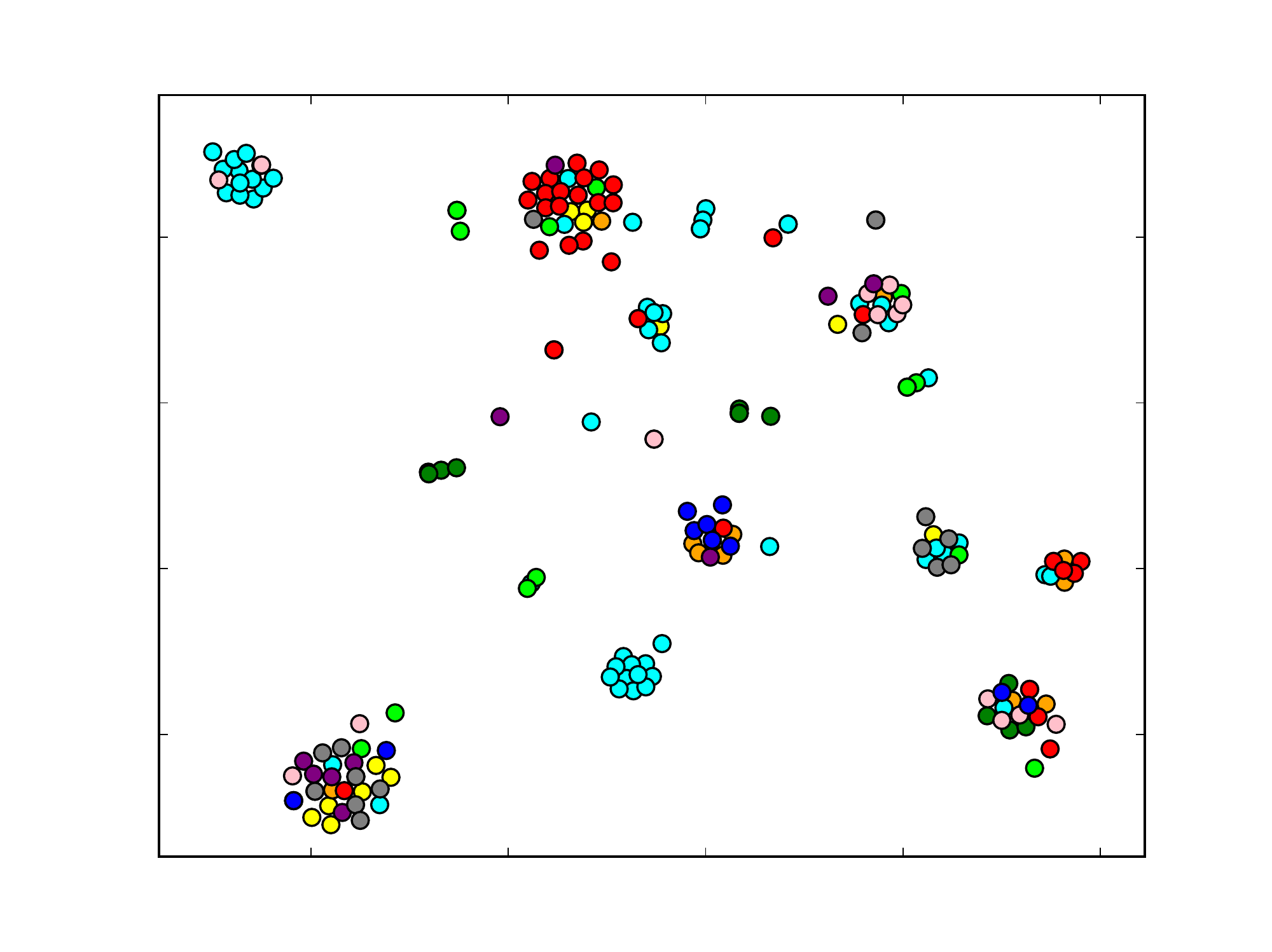}
        \caption{\proposed, 100\% trained.}
        \label{fig:tsne_full}
    \end{subfigure}
    \hfill
    \begin{subfigure}[b]{0.45\columnwidth}
        \centering
        \includegraphics[width=\textwidth, trim={2.5cm 1.5cm 2cm 1.5cm}, clip] {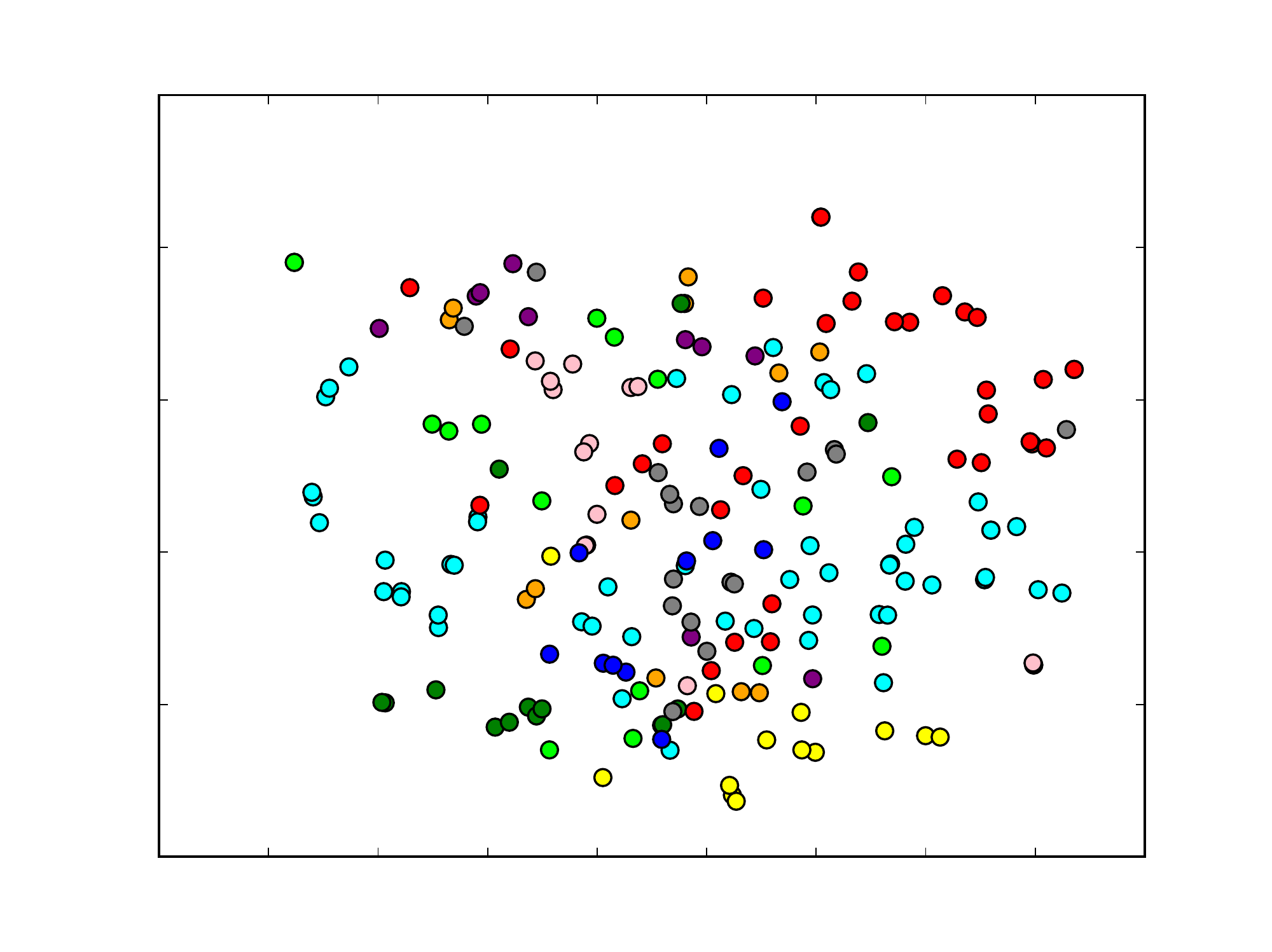}
        \caption{GCE-GNN, 50\% trained.}
    \end{subfigure}
    \hfill
    \begin{subfigure}[b]{0.45\columnwidth}
        \centering
        \includegraphics[width=\textwidth, trim={2.5cm 1.5cm 2cm 1.5cm}, clip] {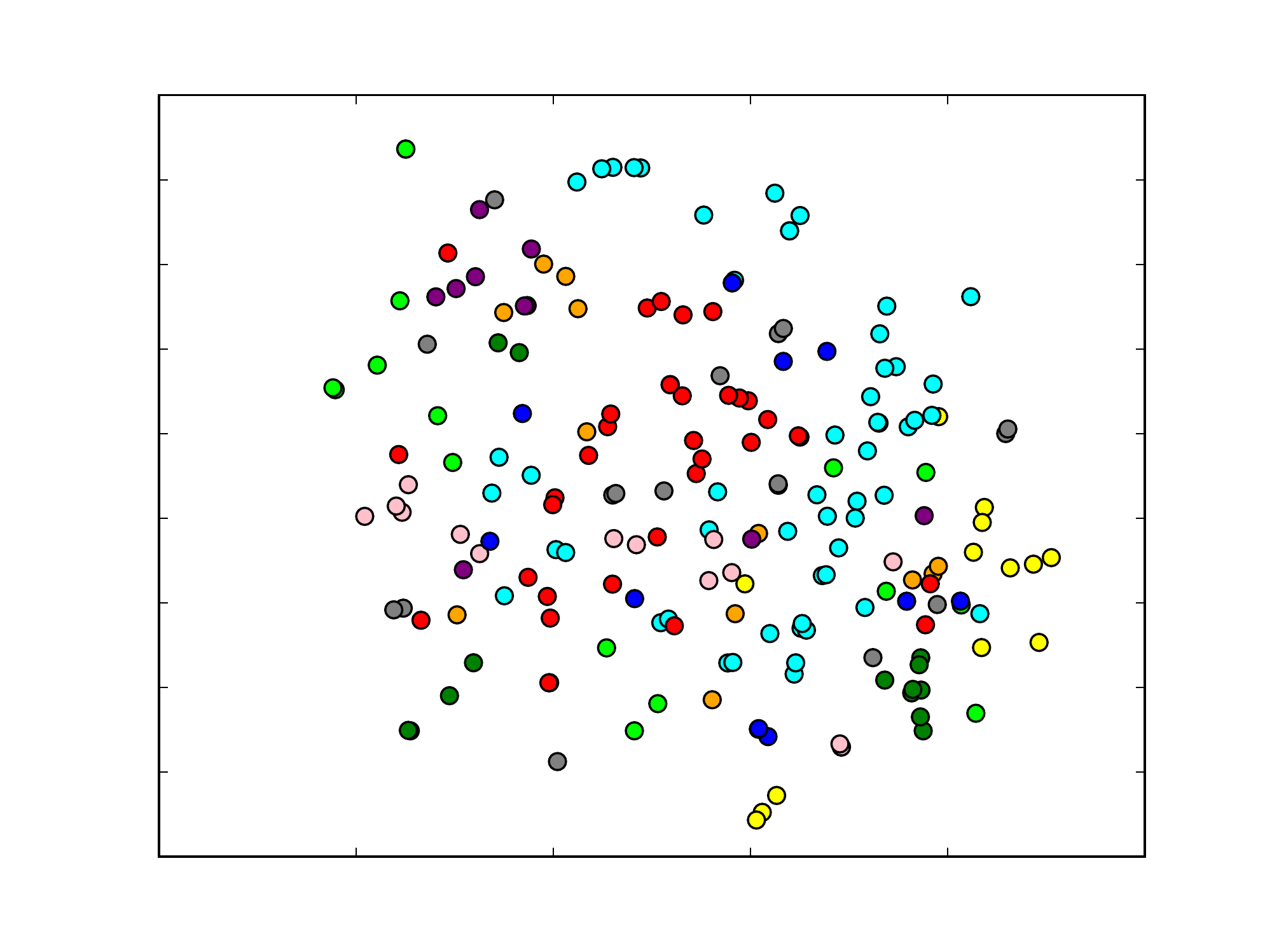}
        \caption{GCE-GNN, 100\% trained.}
    \end{subfigure}
    \hfill
    \begin{subfigure}[b]{0.45\columnwidth}
        \centering
        \includegraphics[width=\textwidth, trim={2.5cm 1.5cm 2cm 1.5cm}, clip]{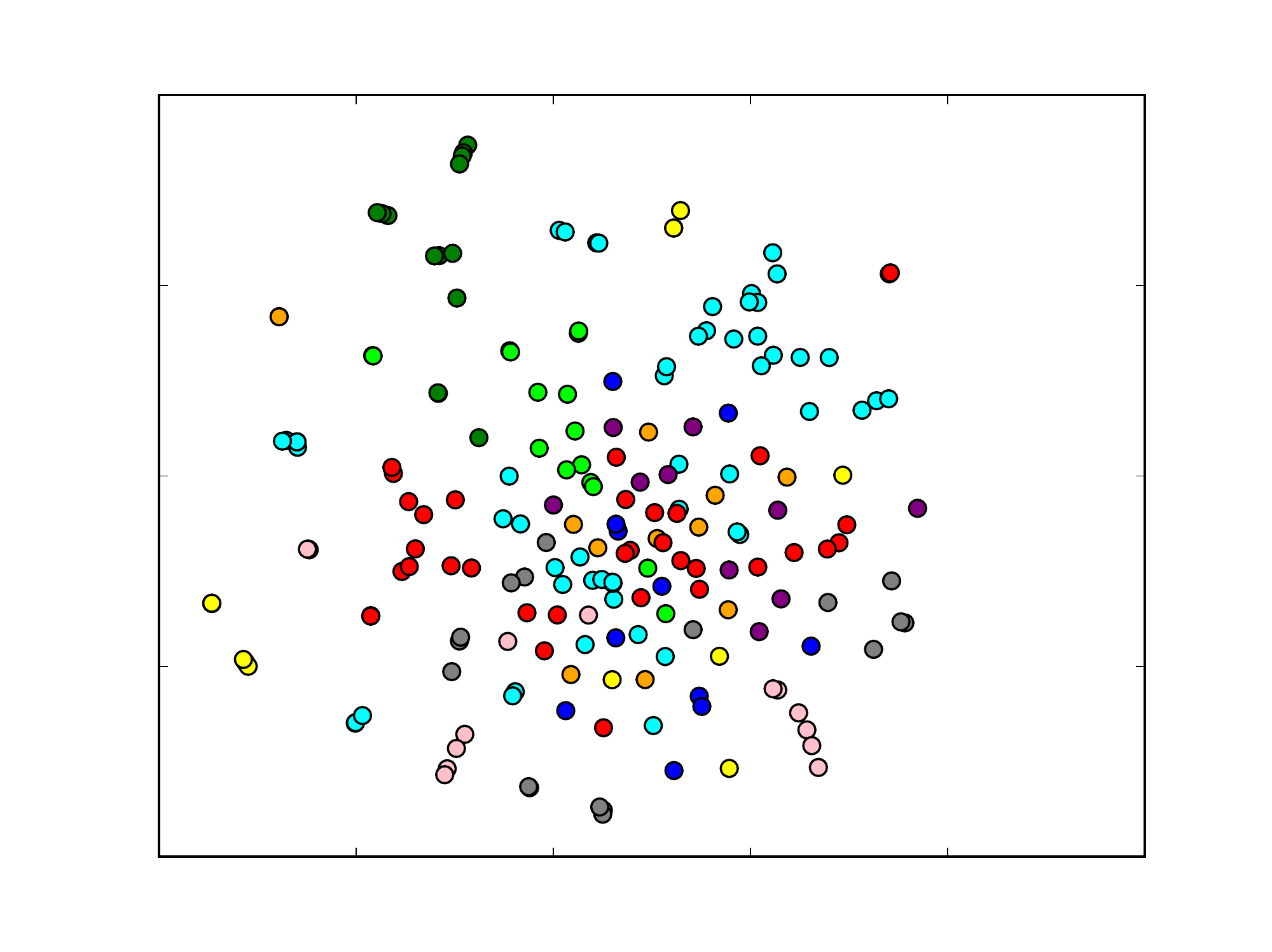}
        \caption{Mean, 50\% trained.}
    \end{subfigure}
    \hfill
    \begin{subfigure}[b]{0.45\columnwidth}
        \centering
        \includegraphics[width=\textwidth, trim={2.5cm 1.5cm 2cm 1.5cm}, clip]{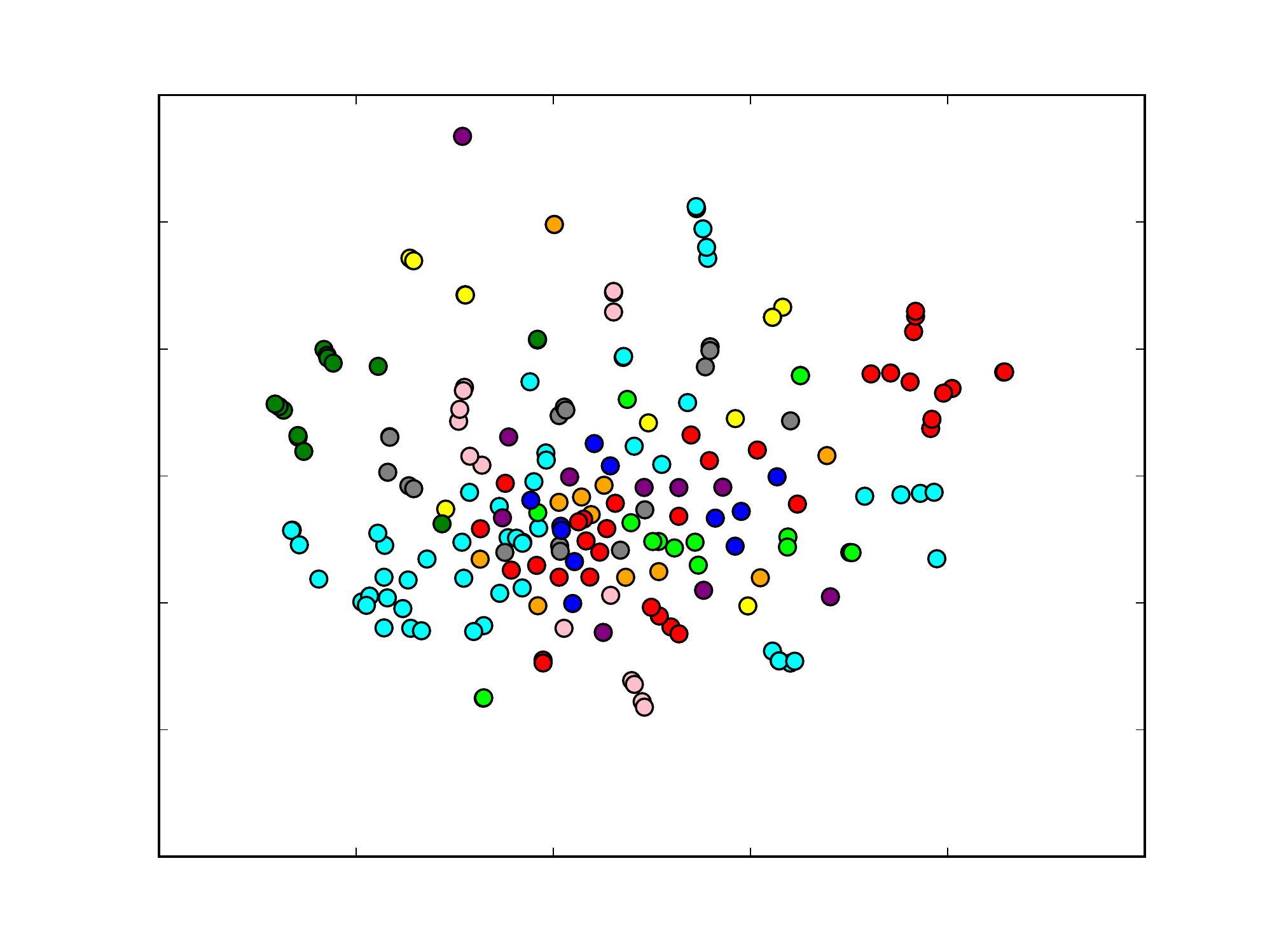}
        \caption{Mean, 100\% trained.}
    \end{subfigure}
    \vspace{-0.2cm}
    \caption{Visualizations of several representations related to sessions in \proposed, GCE-GNN, and the mean encoder (best viewed in color).}
    \label{fig:tsne}
    \vspace{-0.7cm}
\end{figure}

To visually show the superiority of the ability of \proposed to imitate the general interests of users, this section provides visualizations of several latent representations related to the \emph{full} sessions of RetailRocket dataset from \proposed, GCE-GNN, and a simple mean encoder (Fig. \ref{fig:tsne}). The mean encoder is a simple encoder that expresses a session as the mean of the item embeddings within the session, and predicts the next item with the dot product score between the session representation and the item embedding. We used t-distributed Stochastic Neighbor Embedding (t-SNE) \cite{tsne} to visualize the high-dimensional representations. Figure \ref{fig:tsne} demonstrates the representations related to the sessions of 10 random users. The circles of the same color means they belong to the same user. Each circle represents the proxy representation for \proposed, the global feature from the global graph for GCE-GNN \cite{gcegnn}, and the session representation for the mean encoder.

In GCE-GNN and the mean encoder, few representations are classified according to the users. 
In contrast, although there is no user information, the proxies selected by sessions of the same user tend to be clustered.
Some of the proxies are grouped when \proposed is half-trained, and others are grouped as the model is further trained. 
This result further supports our claim: 1) the user's general interest is difficult to be fully captured only by the information within the short session or the information from the neighbor sessions based on item co-occurrence, and 2) the user's general interest can be imitated by a proxy in \proposed which is selected in an unsupervised manner and is shared across several sessions. 

From the visualization of proxies (Fig., \ref{fig:tsne_full}), we can discover a characteristic of proxies in \proposed.
Some proxies are selected by sessions of several users, which is because several users may share similar general interests.
That is, such proxies are associated with universal interests that a number of users have.
Also, there are multiple proxies that are selected by sessions of the same user (e.g., light blue, green, and light green), which means that a proxy can model a more fine-grained information than the general interest of a user, as needed.
As a result, as mentioned above, the proxies in \proposed trained in an unsupervised manner learn information related to the general interest of users by imitating it, but model more detailed information if necessary.

\vspace{-0.2cm}
\section{Conclusion} 
This paper proposes a novel framework for SRS, called \proposed, which uses the input session to select a proxy which imitates the user's general interest in an unsupervised manner, and then predicts the next item of the session considering the proxy and the short-term interest within the session. 
Moreover, we provide a revised version of \proposed for another real-world scenario, where a few sessions have their user information, and achieve a further improvement of recommendation performance on the scenario. 
Through extensive experiments, we show that \proposed considerably outperforms the state-of-the-art competitors by modeling proxies.
Also, our analyses on the proxies in \proposed demonstrate that the information encoded in the proxies actually implies the general interests of users. 

\begin{acks}
This work was supported by the NRF grant funded by the MSIT (No. 2020R1A2B5B03097210), and the IITP grant funded by the MSIT (No. 2018-0-00584, 2019-0-01906).
\end{acks}

\bibliographystyle{ACM-Reference-Format}
\bibliography{reference}

\end{document}